\documentclass{aa}

\usepackage{soul}
\usepackage{graphicx}
\usepackage{color}
\usepackage{txfonts}
\usepackage{natbib}
\usepackage{bbold}
\usepackage{subcaption}
\usepackage{amsmath}
\usepackage{amsfonts,amssymb}

\usepackage[T1]{fontenc}

\usepackage[overload]{empheq}

\usepackage{mwe}

\usepackage[colorlinks=true,linkcolor=black,urlcolor=black,citecolor=black,breaklinks=true]{hyperref}

\begin{document}

    \title{A new multifluid method for dusty astrophysical flows}
    \subtitle{Application to turbulent protostellar collapses}
    
   \author{G. Verrier$^{1}$
   \and U. Lebreuilly$^{2}$ \and P. Hennebelle$^{2}$
}
   \institute{$^{1}$Universit\'{e} Paris Cité, Universit\'{e} Paris-Saclay, CEA, CNRS, AIM, F-91191, Gif-sur-Yvette, France
\email{gabriel.verrier@cea.fr}    \\
$^{2}$Universit\'{e} Paris-Saclay, Universit\'{e} Paris Cité, CEA, CNRS, AIM, 91191, Gif-sur-Yvette, France  
}

   \date{Received March 20, 2025; accepted July 11, 2025}

  \abstract{Stars and planets form in collapsing clouds of gas and dust. The presence of dust grains and their local distribution play a significant role throughout the protostellar sequence, from the thermodynamics and the chemistry of molecular clouds to the opacity of collapsing protostellar cores and the coupling between the gas and the magnetic field and down to planet formation in young and evolved disks.
  }{We aim to simulate the dynamics of the dust, considering the whole range of grain sizes, from few nanometers to millimeters.
  }{We implemented a neutral pressureless multifluid that samples the dust size distribution in the RAMSES code. This multifluid is dynamically coupled to the gas via a drag source term and self-gravity, relying on the Eulerian approach.}{We designed a Riemann solver for the gas and dust mixture that prevents unphysical dust-to-gas ratio variations for well-coupled grains. We illustrated the capacities of the code by performing simulations of a protostellar collapse down to the formation of a first hydrostatic core, both for small and large dust grains. Grains over 100 microns significantly decouple from the gas. The spatial maps and the probability density functions indicate that dust enrichment within the first hydrostatic core and in some locations of the envelope increases as a function of the grain size and the level of initial turbulence.}{
  Thanks to the novel Riemann solver, we recovered the terminal velocity regime, even at low resolution. Moreover, we successfully extended it to regimes where the grain inertia matters. The multifluid module performs the coupling between the dust and the gas self-consistently all through the dynamical scales. The dust enrichment in the first hydrostatic core and the envelope have been revised here, assuming the initial turbulence and grain sizes. This enables us to probe new potential locations, epochs, and initial conditions for planet formation.}
      \authorrunning{Verrier, G., Lebreuilly, U., and Hennebelle, P.}
  \maketitle

\nolinenumbers %remove linenumbers

\section{Introduction}

Dust grains provide the solid material to form planets. However, the questions of how and when the first planetesimals form are still a matter of discussion \citep{2023ASPC..534..717D}. Star formation and planetary formation are intimately connected by dust evolution. Even though dust represents a small fraction of the mass (about 1 percent), its dynamics strongly affects the coupling of the bulk mass (i.e., the gas) to the magnetic field, which is a prime mechanism in regulating angular momentum during a protostellar collapse \citep{2022FrASS...9.9223M,2023ASPC..534..317T}. Indeed, dust grains carry most of the charges and, thus, they control the magnetic resistivities \citep{2016A&A...592A..18M,2016MNRAS.460.2050Z}. Moreover, because the dust opacity mainly controls the optical thickness, it determines how much energy is radiated away during the collapse \citep{1963ApJ...138.1050G,1969MNRAS.145..271L}. Finally, dust grains allow for recombination and chemistry on their surfaces, making them a preferential channel for formation for abundant molecules such as H$_{\rm{2}}$ \citep{1963ApJ...138..393G}.

The multiple roles of the dust depend on the local size distribution and dust-to-gas ratio, which define, for instance, the available surface for chemistry and the mass for dynamical instabilities, such as the polydisperse streaming instability \citep{Krapp2019}. Mathis, Rumpl, and Nordsieck (hereafter MRN) found, by fitting the interstellar extinction, a power-law size distribution ranging from 5 nm to 250 nm \citep{1977MRN}, whereby large grains carry most of the dust mass and the small grains vastly outnumber large grains and provide most of the surface. Such conclusions rely on the optical and thermal properties of dust grains, resulting from the grain size, structure (compact or aggregate), and composition (grains are thought to be mainly carbonaceous and silicated). Efforts have to been made to better identify and trace the dust distribution, for example, by including the spectral energy distribution from Herschel and Planck \citep{2011A&A...525A.103C} as well as laboratory measurements \citep{2024A&A...684A..34Y}. However, it remains poorly constrained during stellar formation. 

In most astrophysical fluid codes, dust is accounted for in the bulk mass. We can assume, for instance, that the gas and dust mixture is perfectly coupled with constant dust-to-gas ratio and dust distribution to compute the opacity and the evolution of the radiation field \citep{2023A&A...678A.162M} or the evolution of the magnetic field via magnetic resistivities \citep{2016MNRAS.457.1037W}. 
The interest in the dust dynamics in the context of protoplanetary disk evolution and planet formation has greatly
motivated dedicated numerical developments. The most frequently used codes include PHANTOM \citep{2017ascl.soft09002P}, FARGO3D \citep{2016ApJS..223...11B}, Athena++ \citep{2020ApJS..249....4S}, and Idefix \citep{2023A&A...677A...9L}. They can rely on different approaches such as Lagrangian particles in SPH codes as implemented in PHANTOM, or as a multifluid in grid-based codes as implemented in FARGO3D \citep{2019ApJS..241...25B} and Athena++ \citep{2022ApJS..262...11H}. More specifically, \citet{2019ApJS..241...25B} implemented a first-order implicit drag scheme whose direct inversion for noncolliding dust species is presented in \citet{2020RNAAS...4..198K}. Subsequently, \citet{2022ApJS..262...11H} implemented two second-order schemes that remain stable in stiff regimes. Developing higher-order schemes coupling the hydrodynamics and the drag source term is an active area of research \citep{2023A&A...673A..66K,2024ApJS..271....7K}. Moreover, it can benefit high-performance from adaptive mesh refinement (AMR) in \citet{2020ApJS..249....4S, 2023A&A...673A..66K} and GPU parallelization in \citet{2016ApJS..223...11B,2023A&A...677A...9L}.

In this work, we address the implementation of a multifluid solver in the RAMSES code \citep{RAMSES}. RAMSES is a 3D finite-volume AMR code. The adaptive mesh refinement (AMR), together with subcycling, are essential to performing stellar formation simulations. Moreover, many physics modules have been successively implemented. Among the most useful ones in star formation simulations are radiative transfer \citep{2011A&A...529A..35C,2015MNRAS.449.4380R,2015A&A...578A..12G}, ideal magnetohydrodynamics \citep{2006A&A...457..371F}, and standard non-ideal magnetohydrodynamics (MHD), namely, Ohm diffusion and ambipolar drift \citep{2012ApJS..201...24M}, as well as the Hall effect \citep{2018A&A...619A..37M}. For these reasons, RAMSES is a powerful tool to deal with the whole range of densities and scales in which dust plays a role (see, e.g., \citet{2023A&A...680A..23A} for the dynamical range to reach the birth of a protostar from a protostellar collapse). In particular, this could capture multi-scale and multi-physics processes such as infall, turbulence, and other transport mechanisms. 

The first implementation of the dust dynamics in RAMSES was done by \citet{2019A&A...626A..96L}. The dust dynamics is computed in the terminal velocity approximation, whose validity fails for strongly decoupled grains, which is likely to occur for (dynamically) large grains or charged grains. Moreover, this approximation fails in shocks \citep{2019MNRAS.488.5290L}.

Recently, charged dust has been introduced as massive superparticles (particle-in-cell method) coupled to the ideal MHD equations \citep{2023MNRAS.518.2825M}.

The paper is organized as follows. We present the methods in Sect. \ref{sec:methods} and the validation tests in Sect. \ref{sec:validation_tests}. In particular, we include tests with multiple dust species and various coupling regimes. We compare to the terminal velocity approximation in protostellar collapses of dense cores and assess the limitations of the approaches in Sects. \ref{sec:collapse_small} and \ref{sec:collapse_large}. In Sect. \ref{sec:collapse_turb}, we evaluate dust enrichment in the first hydrostatic core and the envelope as a function of the grain size and the initial turbulence. Section \ref{sec:conclusion} presents our conclusions. More details on the numerical results are given in the appendix.

\section{Methods} \label{sec:methods}

\subsection{Equations} \label{sec:equations}

We consider a multifluid of $\mathcal{N}$ dust species, each one referred by the index $d$:
\begin{equation}
    \partial_t \rho_d + \nabla \cdot (\rho_d \mathbf{V}_d ) = 0,
    \label{eq:dust_density_conservation}
\end{equation}
\begin{equation}
    \partial_t (\rho_d \mathbf{V}_d) + \nabla \cdot \left(  \rho_d \mathbf{V}_d \mathbf{V}_d \right) = -\rho_d \nabla \phi +   \mathbf{f}_{g \to d}.
    \label{eq:dust_momentum_conservation}
\end{equation}
Each dust species is coupled to the gas via an individual drag force $\mathbf{f}_{g \to d}$. Then, $\phi$ is the gravitational potential. Contrary to the dust fluids, the gas is pressure-supported and, thus, the gas energy equation is required. The equations of the gas dynamics are
\begin{equation}
    \partial_t \rho_g + \nabla \cdot (\rho_g \mathbf{V}_g ) = 0,
    \label{eq:gas_density_conservation}
\end{equation}
\begin{equation}
    \partial_t (\rho_g \mathbf{V}_g) + \nabla \cdot (  \rho_g \mathbf{V}_g \mathbf{V}_g + P_g \mathbb{1} 
    ) = -\rho_g \nabla \phi + \sum_d  \mathbf{f}_{d \to g},
    \label{eq:gas_momentum_conservation}
\end{equation}
\begin{equation}
    \partial_t E_g + \nabla \cdot \left( (E_g +P_g 
    ) \mathbf{V}_g    \right) = -\rho_g \nabla \phi \cdot \mathbf{V}_g + \sum_d \mathbf{f}_{d \to g} \cdot \mathbf{V}_g + Q.
    \label{eq:gas_energy_conservation}
\end{equation}

Each dust fluid back-reacts on the gas thus $\mathbf{f}_{g \to d} = - \mathbf{f}_{d \to g}  $. We can express the drag force as 
\begin{equation}
    \mathbf{f}_{g \to d} = \frac{\rho_d }{t_{s,d}} ( \mathbf{V}_g - \mathbf{V}_d),
\end{equation} defining $t_{s,d}$ as the stopping time of the dust grain in the gas fluid. In the Epstein regime \citep{1924PhRv...23..710E}, it is
\begin{equation}
    t_{s,d}=\sqrt{\frac{\pi \gamma}{8}} \frac{\rho_{\rm{grain,}d}}{\rho_g} \frac{s_{\rm{grain,}d}}{c_s},
    \label{eq:eipstein_stopping_time}
\end{equation}
where $c_s$ is the
sound speed of the gas, $\gamma$ its adiabatic index and $s_{\rm{grain,}d}$ is the size (radius) of the grain, while $\rho_{\rm{grain,}d}$ its intrinsic density.

Here, $E_g$ and $P_g$ are the total energy and the pressure of the gas, respectively, while $Q$ is the energy deposit in the gas due to frictional heating. When it is not set to zero, to conserve the energy of the gas-dust multifluid system, we can express it as

\begin{equation}
    Q = - \sum _d\left( \mathbf{f}_{g \to d} \cdot \mathbf{V}_d +  \mathbf{f}_{d \to g} \cdot \mathbf{V}_g \right) = \sum_d \frac{\rho_d }{t_{s,d}}   (\mathbf{V}_g - \mathbf{V}_d)^2 .
\end{equation}

Finally, the gravitational potential, which both the gas and the dust multifluid contribute to, can be obtained using the Poisson equation
\begin{equation}
    \Delta \phi = 4 \pi G \left(\rho_g + \sum_d \rho_d \right).
\end{equation}

\subsection{Operator splitting}

Equations \eqref{eq:dust_density_conservation}-\eqref{eq:gas_energy_conservation} can be written as follows:
\begin{equation}
    \partial_t \mathbf{U}_g + \mathbf{\nabla} \cdot \mathbf{F}_g (\mathbf{U}_g) = \mathbf{S}_{\rm{grav},g} + \mathbf{S}_{\rm{drag},g},
    \label{eq:operator_splitting_gas}
\end{equation}
\begin{equation}
    \partial_t \mathbf{U}_d + \mathbf{\nabla} \cdot \mathbf{F}_d (\mathbf{U}_d) = \mathbf{S}_{\rm{grav},d} + \mathbf{S}_{\rm{drag},d}, d \in \{1,.. \mathcal{N} \},
    \label{eq:operator_splitting_dust}
\end{equation}
where $\mathbf{U}_g$ and $(\mathbf{U}_d)_{d \in \{1,.. \mathcal{N} \}}$ are the conserved variables of the gas and of the dust multifluid. Then, $\mathbf{F}_g$ and $(\mathbf{F}_d)_{d \in \{1,.. \mathcal{N} \}}$ are the fluxes corresponding to the hydrodynamics of the conserved variables, $\mathbf{S}_{\rm{grav},g}$ and $(\mathbf{S}_{\rm{grav},d})_{d \in \{1,.. \mathcal{N} \}}$ the gravity source terms, and $\mathbf{S}_{\rm{drag},g}$ and $(\mathbf{S}_{\rm{drag},d})_{d \in \{1,.. \mathcal{N} \}}$ the drag source terms.

The hydrodynamical scheme and the gravity source term are already implemented in RAMSES \citep{RAMSES}. We include the drag source term using a Lie-Trotter splitting, as explained hereafter, carrying out the computation of the whole step over $\Delta t$ as two successive steps over $\Delta t$. 

The hydrodynamical step is performed by an unsplit second-order Godunov integrator using the MUSCL-HANCOCK scheme with various slope limiters, including minmod \citep{1986AnRFM..18..337R}, superbee \citep{1986AnRFM..18..337R}, and Van-Leer \citep{1974JCoPh..14..361V}. When using multiple levels of mesh refinement, multifluid variables can be restricted from fine levels to coarse levels by averaging down and they can be prolongated from coarse levels to fine levels using different linear interpolation strategies (minmod slope, monotonized central slope, unlimited central slope). This prolongation can be applied either to the conservative variables or to the primitive variables, which are $(\rho_d, \rho_d \mathbf{V}_d)$ and $(\rho_d,\mathbf{V}_d)$ respectively. Various Riemann solvers are implemented, as described in the next section. The gravity source term is added to the hydrodynamical term, following a second-order midpoint scheme.

After computing the hydrodynamical and gravity steps, the drag step is performed using the first-order implicit scheme addressed in \citet{2020RNAAS...4..198K} on the velocity of the gas and the dust species. In the absence of frictional heating, the kinetic energy of the gas is simply updated. To account for the frictional heating, we remove the increment of the kinetic energy in the dust multifluid to the total energy of the gas. We test the order of the scheme in Sect. \ref{sec:dustywave}.

\subsection{Riemann solver for a gas and dust multifluid mixture} \label{sec:HLLgd}

The Riemann solver of the multifluid requires special care, depending on the coupling of the dust with the gas. Usually, multifluid Riemann solvers split into independent problems for each fluid. This strategy is entirely valid in the absence of interactions between the fluids. However, when dealing with infinitely low Stokes grains (dynamically small grains), we could not recover dust-to-gas ratio variations vanishing to zero (see Sect. \ref{sec:collapse_small}), while the drag solver was correctly adapting the dust velocity to the one of the gas. Therefore, this numerical difference only originates from the advection of the dust density, which comes from the computation of Riemann fluxes. As far as we know, this problem and alternative strategies, such as using individual Riemann solvers or one common Riemann solver, have not been discussed in previous dust multifluid implementations. Our new solving strategy for interacting gas and dust is based on the propagation of waves in the mixture.

In the absence of source terms, for a pressureless dust fluid, the Riemann problem leads to solutions of delta shocks and vacuum states \citep{LeVeque2004}. We implemented the Riemann solver of \citet{2022ApJS..262...11H}, based on upwind fluxes, as a reference solver used in other codes. We implemented a local Lax-Friedrichs (LLF) solver, denoted as LLFd and presented in Appendix \ref{sec:LLFd}. 

When the dust is well coupled to the gas, the mixture behaves as a single fluid and, thus, the same waves propagate in the dust and in the gas. In the Riemann solver presented hereafter, we consider the same wave fan for the gas and for the dust fluid, based on the HLL approach, to model this strong coupling situation. It can switch to the individual local Lax-Friedrichs solver for a specific dust fluid if a specific decoupling criterion is met. Such a criterion is based on the dust fluid kinematics and its purpose is to inject the right amount of numerical viscosity depending on the situation. Considering more appropriate wave speeds in the Riemann solver should stabilize the hydrodynamical scheme.

We go on to define the solvers denoted by HLLgd and LLFgd (when the HLL solver or the LLF solver, respectively, is used for the gas). For the Riemann problem at the interface $i+1/2$, we consider the left and right states of the dust multifluid, $U_{d,L}$ and $U_{d,R}$. The normal components of the dust velocity are $V_{d,L}$ and $V_{d,R}$ and the fluxes are denoted by $F_{d,L}= F_d(U_{d,L})$ and $F_{d,R}= F_d(U_{d,R})$. The Riemann problem for the gas, neglecting the interaction with the multifluid, is modeled by the two wave speeds, $S_{g,L}$ and $S_{g,R}$, for the HLL solver \citep{articleHLL}, such that $S_{g,L}<S_{g,R}$. We define the flux at the interface for each dust fluid, $F_{d,i+1/2}$, as follows.\\

If $S_{g,L}>0$ (upwind wave speeds of the gas from the left), then
\begin{equation}
F_{d,i+1/2} =\begin{cases}
   F_{d,L} & \text{if } V_{d,L}, V_{d,R}>0 \text{ (dust upwind flow)}, \\
   F_{LLF,d}  & \text{otherwise (switch).}
\end{cases}
\end{equation}

Symmetrically, if $S_{g,R}<0$ (upwind wave speeds of the gas from the right), then
\begin{equation}
F_{d,i+1/2} =\begin{cases}
   F_{d,R} & \text{if } V_{d,L}, V_{d,R}<0 \text{ (dust upwind flow)} , \\
   F_{LLF,d}  & \text{otherwise (switch).}
\end{cases}
\end{equation}

Otherwise $S_{g,L} \leq 0 \leq S_{g,R}$, the solution at the interface can be modeled as
\begin{equation}
F_{d,i+1/2} =\begin{cases}
   F_{HLL,d}(S_{g,L},S_{g,R}) & \text{if } S_{g,L} < V_{d,L}, V_{d,R}< S_{g,R}, \\
   F_{LLF,d}  & \text{otherwise (switch),}
\end{cases}
\end{equation}
depending on whether the left and right dust velocities belong to the gas wave fan (first subcase) or not (switch).\\

We define the HLL flux as
\begin{equation}
    F_{HLL,d}(s_L, s_R) = \frac{s_R F_{d,L} -s_L F_{d,R} + s_R s_L( U_{d,R} - U_{d,L})  }{s_R - s_L}.
\end{equation}

\begin{figure}
    \resizebox{\hsize}{!}{\includegraphics{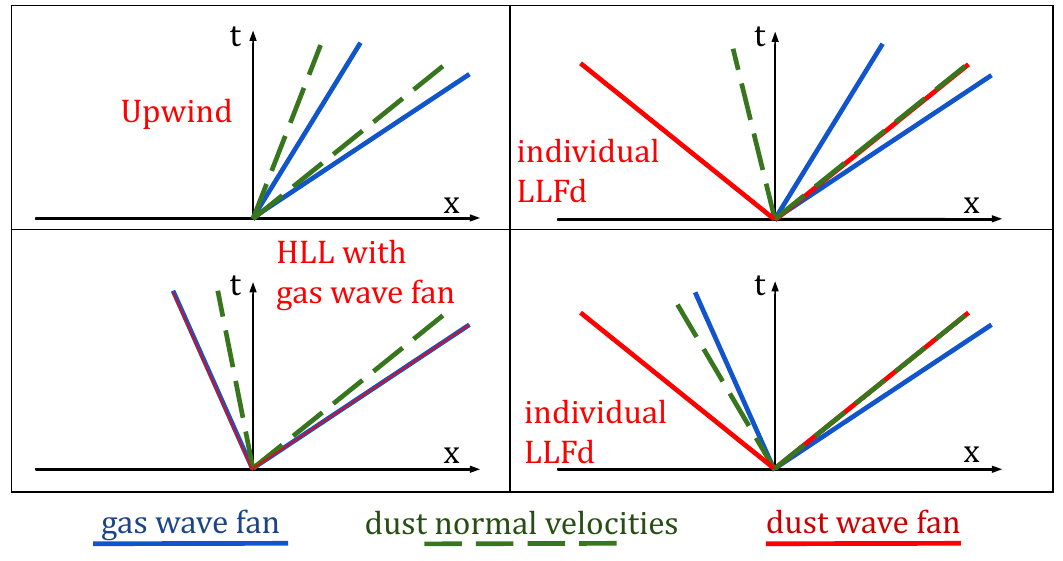}}
    \caption{Modeling of the HLLgd solver of each dust fluid depending on the Riemann problem at the interface: Riemann problem on the gas in lines and kinematic coupling situation in columns. In blue, the gas wave fan defined by $S_{g,L}$ and $S_{g,R}$. With the dotted green lines, we show the dust normal velocities, $V_{d,L}$ and $V_{d,R}$. The HLLgd flux for the dust species $d$ is described in red by its wave fan.}
    \label{fig:HLLgd}
\end{figure}

LLFgd is the particular case of HLLgd where $S_{g,R}=- S_{L,R}>0$. The different Riemann problems with the corresponding solutions of the HLLgd solver are illustrated in Fig. \ref{fig:HLLgd}.

We have decided to switch to a local Lax-Friedrichs solver (LLFd, see Appendix \ref{sec:LLFd})
because of its simplicity, following the HLL approach, and robustness, and also because some situations could be difficult to model properly, for example if $V_{d,L}, V_{d,R}>0$ and $S_{g,R}<0$, which is when the gas and the dust flows cross each other.

We propose an interpretation of the switch criterion in the linear regime. For small perturbations in the gas and dust velocities, $\delta v_g$ and $\delta v_d$, and for a compressible wave propagating at the wave speed $c_\phi$, the equations of mass conservation lead to

\begin{equation}
    \frac{\delta \theta_d}{\theta_d} = \frac{\delta v_d - \delta v_g}{c_\phi},
\end{equation}
where $\theta_d = \rho_d/\rho_g$ is the dust-to-gas ratio. More generally, we could expect a significant variation of the dust-to-gas ratio if $|\delta v_d - \delta v_g|>c_\phi$. This switch criterion also means that the dust species is not in the influence area of the gas, modeled by its wave fan. We note that the wave speed values used for the gas wave fan overestimate the wave speeds expected for a perfect monofluid. Indeed, the usual MHD wave speeds scales with $1/\sqrt{\rho}$, where $\rho =\rho_g + \sum_d \rho_d $ is the bulk mass carrying the wave; for example, the effective sound speed is $c_{s,\rm{eff}}= c_s/\sqrt{1+\sum_d \rho_d/\rho_g}<c_s$ \citep{2011MNRAS.418.1491L}. This overestimation is weak for a small dust-to-gas ratio and it provides a small and safe boost of numerical viscosity.

As far as we know, adapting the Riemann solver of the multifluid to model the dynamical coupling with the gas has never been considered and thus we have evaluated the impact of the choice of the Riemann solver for studying the dust decoupling in the conditions of protostellar collapses (Sect. \ref{sec:protostellar_collapses}). The source terms are not explicitly included in the approximation of the solution of the Riemann problem. Jump conditions on fluxes are no longer satisfied and thus the development of an extension of the HLLgd method to a HLLD wave fan, following \citet{2005JCoPh.208..315M}, would require a dedicated work. Because fractional step schemes can fail in capturing states that are close to equilibrium, \citet{LeVeque1999} incorporate the source terms in the Riemann problem. However, this approach applies to the specific situation of quasi-steady states, which is not likely to occur for a vast range of grain sizes. Developing a more general and sophisticated Riemann solver is beyond the scope of this work. 

\subsection{Time-stepping}

One advantage of the implemented implicit drag scheme is to be asymptotically and unconditionally stable \citep{2020RNAAS...4..198K}. An explicit scheme would require to limit the time step to the stopping times of the dust species, which could be problematic during a protostellar collapse simulation for which the stopping times vary a lot because of the size distribution and the local density.

The Courant-Friedrichs-Lewy (CFL) condition \citep{1928MatAn.100...32C} should guarantee that any information cannot leave the cell during one time step. We therefore add the classical condition for each dust species $d$ and for each cell sharing a common subcycled domain,

\begin{equation}
    \Delta t < C_{\rm{CFL}} \frac{\Delta x}{ \max_j (|(V_{d,x})_j| + | (V_{d,y})_j|  +| (V_{d,z})_j)|) },
\end{equation}
where $C_{\rm{CFL}}<1$ is a safety factor. In the case of a highly coupled mixture, the sound speed of the monofluid should be considered. This should not be problematic because the CFL condition on the hydrodynamical step of the gas should be sufficient, as discussed in Sect. \ref{sec:HLLgd}. We have not found it necessary to include the acceleration due the drag force because this force tends to reduce the velocity drifts between the fluids.

\section{Validation tests} \label{sec:validation_tests}

We test the coupling between the hydrodynamical scheme, the drag solver, and the self-gravity solver. We test the drag solver alone with the dustybox test (Sect. \ref{sec:dustybox}) and its coupling with the hydrodynamical solver with the dustywave test (Sect. \ref{sec:dustywave}). Both tests were developed in \citet{2011MNRAS.418.1491L} and tested in \citet{2019ApJS..241...25B}. We go further in the dustywave test, by checking the time convergence (Sect. \ref{sec:dustywave_convergence}), the static mesh refinement, and the time subcycling (Appendix \ref{sec:dustywave_amr}). We test the self-gravity with the dustyjeanswave test (Appendix \ref{sec:dustyjeanswave}), similarly to the test in \citet{2024ApJS..271....7K}. We recover the damping mode in a distribution of five dust fluids. We perform the disk settling test with ten dust fluids (Sect. \ref{sec:settling}), as in \citet{2018MNRAS.476.2186H}, and the shock test (Appendix \ref{sec:dustyshock}), following \citet{2019ApJS..241...25B}.\\

In the dustybox, dustywave, and dustyjeanswave tests, the initial conditions are
\begin{equation}
    \rho_g(x,0) = (\rho_g)_0 + \delta (\rho_g)_0 \sin( kx+\phi_g), 
\end{equation}
\begin{equation}
V_g(x,0) = (v_g)_0 \sin(kx+\psi_g).
\end{equation}
A similar expression stands for the dust species. The tests are isothermal, setting $P_g = C_s^2 \rho_g$.
The numerical domain is a periodic 1D box of a length $L=1$. We denote $M_x= L/\Delta x = 2^l$ as the number of cells of a uniform grid of refinement level $l$ and $n=t/\Delta t$ as the number of time steps at the coarsest level of mesh refinement.

\subsection{Dustybox} \label{sec:dustybox}

\begin{figure}
    \resizebox{\hsize}{!}{\includegraphics{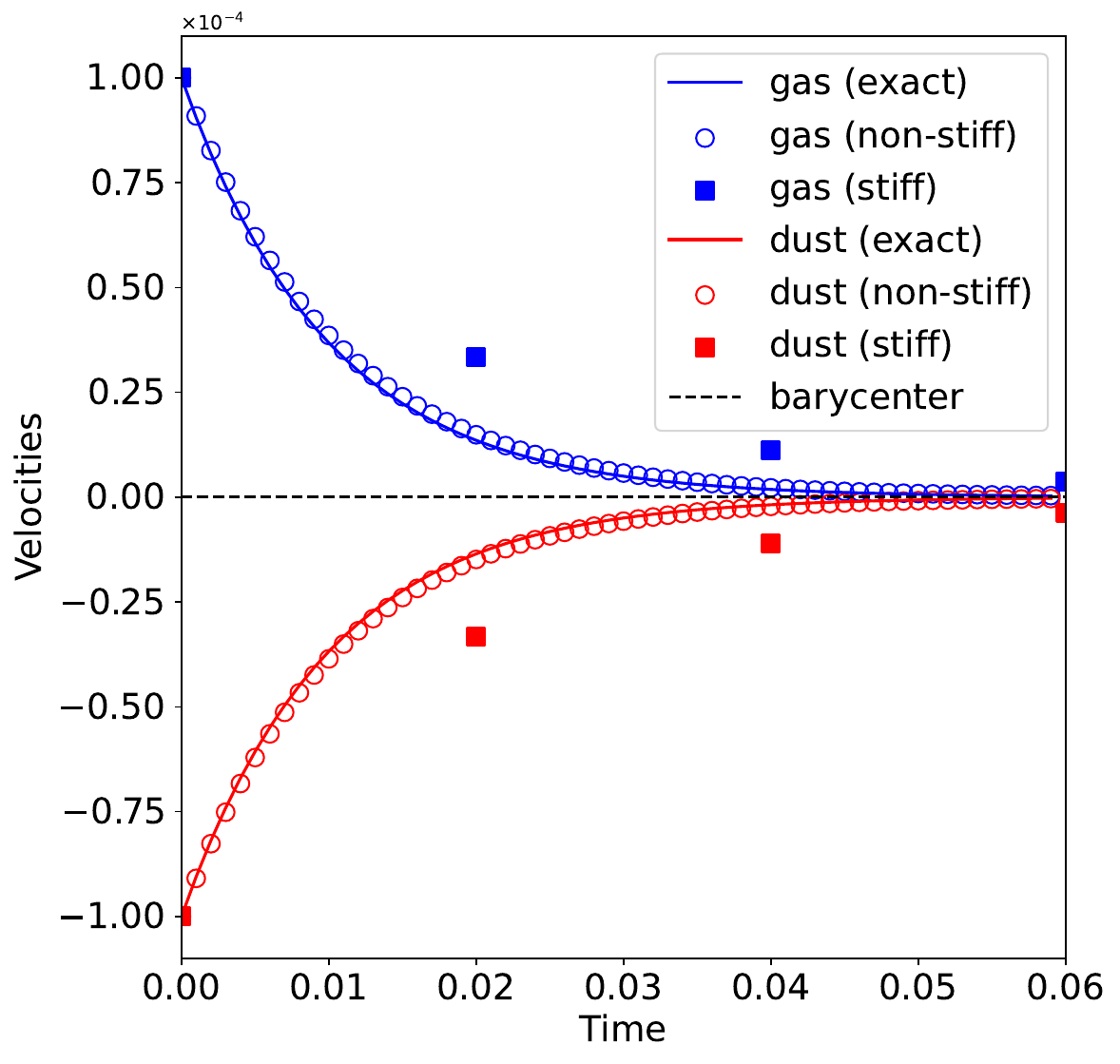}}
    \caption{Exact solution (solid lines) and two numerical solutions (circles and squares, respectively) of the dustybox at $x=0.25$, relaxing towards the barycenter velocity (dotted line) over six damping times ($t=0.06$). Equivalently, $t\approx 3t_{s,d}$, where $t_{s,d}=\rho_d/K_d \approx 0.02$ is the stopping time. Numerical solutions of the first-order implicit drag solver are computed in the stiff regime ($\Delta t = t_{s,d}=0.02$, squares) and in the non-stiff regime ($\Delta t =0.05 t_{s,d} = 10^{-3}$, circles).
    }
    \label{fig:dustybox_relaxation}
\end{figure}

The dustybox test consists in the relaxation towards the velocity barycenter of colliding fluids. It tests the drag solver only. In other words, the flux terms and the gravity terms in Eqs. \eqref{eq:operator_splitting_gas} and \eqref{eq:operator_splitting_dust} are ignored. The evolution of the conservative variables or, more precisely, the momentum of the fluids, are governed by the drag force, expressed as
\begin{equation}
    \mathbf{f}_{g \to d} = K_d( \mathbf{V}_g - \mathbf{V}_d),
    \label{eq:drag_force_dustywave}
\end{equation} 
where $K_d$ is a constant drag coefficient. For two fluids, the evolution of the state $U=(\rho_g v_g, \rho_d v_d)^T$ is given by the system

\begin{equation}
\partial_t 
    \begin{pmatrix}
        \rho_g(x) v_g(x,t) \\
        \rho_d(x) v_d(x,t)
    \end{pmatrix}  = M(x) 
    \begin{pmatrix}
        \rho_g(x) v_g(x,t) \\
        \rho_d(x) v_d(x,t)
    \end{pmatrix}
    ,\label{eq:dustybox_system}
\end{equation}

\begin{equation}
    M(x)=K_d
    \begin{pmatrix}
        -1/\rho_g(x) & 1/\rho_d(x) \\
        1/\rho_g(x) & -1/\rho_d(x)
    \end{pmatrix}.
\end{equation}
The drag source term is colocated with the conservative variables, so the system \eqref{eq:dustybox_system} to solve is an ordinary differential equation (ODE) in time. The exact solution is
\begin{equation}
    U(x,t)=\exp(M(x)t)U(x,0).
    \label{eq:drag_exponential}
\end{equation}
For only one dust species, it is easy to compute explicitly \citep{2011MNRAS.418.1491L}. Therefore, Eq. \eqref{eq:drag_exponential} provides, over one time step $\Delta t$, an exact drag solver and we will refer to it later.

We tested the solver with the initial parameters $K_d=50$, $k=2 \pi$, $(\rho_g)_0=(\rho_d)_0=1$, $\delta (\rho_g)_0=\delta (\rho_d)_0=10^{-4}$, $(v_g)_0=(v_d)_0=10^{-4}$, $\phi_g = \phi_d = 0$, $\psi_g = 0$, and $\psi_d = - \pi$. We recovered the relaxation of gas and dust velocities towards the barycenter, even in stiff regimes (Fig. \ref{fig:dustybox_relaxation}),
in agreement with the original work \citep{2019ApJS..241...25B}. In Appendix \ref{sec:dustybox_sequence}, we go further by proving that 
the drag solver computes the time sequence of the first-order implicit Euler scheme and we present the solution within the entire box domain.

\subsection{Dustywave} \label{sec:dustywave}

\begin{figure}
    \resizebox{\hsize}{!}{\includegraphics{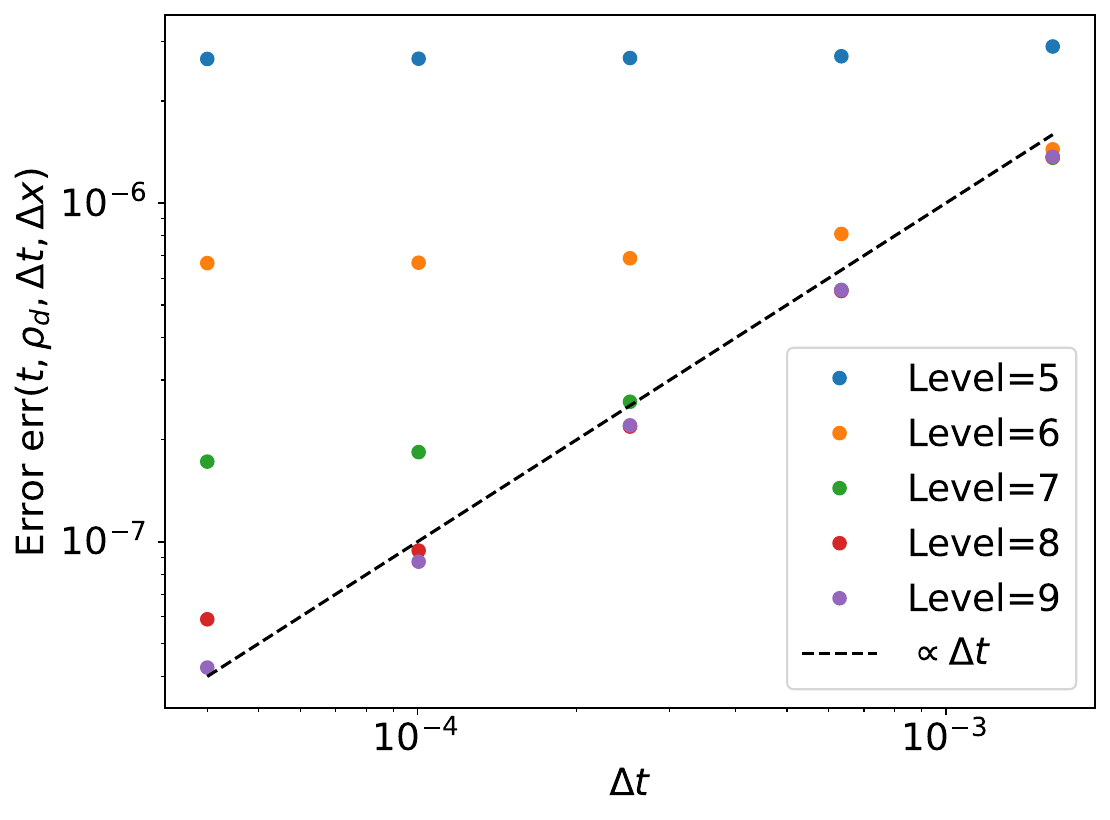}}
    \resizebox{\hsize}{!}{\includegraphics{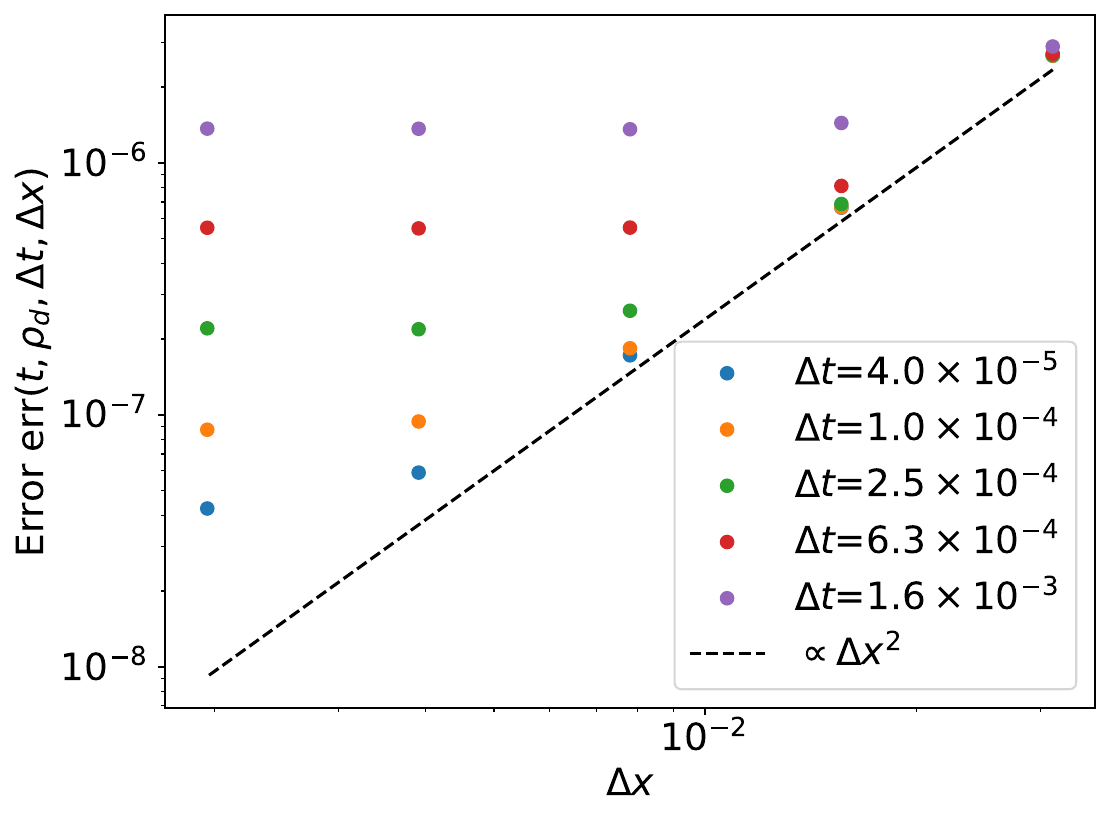}}
    \caption{Error (in time: upper panel and in space: lower panel) of the global scheme on the dust density variable with individual local Lax-Friedrichs solvers on the gas and on the dust (LLFg-LLFd). Test in the strong coupling regime (the setup parameters are the same as for the dustybox test (Sect. \ref{sec:dustybox}), expecting $ \psi_g = \psi_d = 0$).}
    \label{fig:dustywave_errors}
\end{figure}

\subsubsection{Linearized equations} 

We test the scheme resulting from the coupling between the drag solver to the hydrodynamical solver. The linearization of the equations Eqs. \eqref{eq:dust_density_conservation}-\eqref{eq:gas_momentum_conservation} provides the evolution of the perturbations:

\begin{equation}
    d_t 
    \begin{pmatrix}
        \delta \rho_g \\
        \delta \rho_d \\
         v_g \\
         v_d 
    \end{pmatrix}
    =
    \begin{pmatrix}
        0 & 0 & -ik \rho_g & 0 \\
        0 & 0 & 0 & -ik \rho_d \\
        -ik C_s^2/\rho_g  & 0 & -K_d/\rho_g & K_d/\rho_g \\
        0 & 0 & K_d/\rho_d & -K_d/\rho_d
    \end{pmatrix}
        \begin{pmatrix}
        \delta \rho_g \\
        \delta \rho_d \\
         v_g \\
        v_d 
    \end{pmatrix}
    .\label{eq:dustywave_pertubation_system}
\end{equation}

 \subsubsection{Convergence} \label{sec:dustywave_convergence}

We found it necessary to test the convergence of the resulting scheme both in space and in time. Indeed, we identify potential sources of spatial error when coupling the hydrodynamical solver with the drag solver in Appendix \ref{app:FV_drag}. Moreover, we check the error due to the operator splitting (in time). To do so, we define the convergence error at time $t$ on the field $U$ as
\begin{equation}
    \mathrm{err}(t,U,\Delta t, \Delta x) = \frac{1}{M_x} \sum_{k \in [1,M_x]} |U_{\rm{num},k}^n - U_{\rm{ref}}(x_{k+1/2},t)|.
    \label{eq:convergence_error}
\end{equation}

Here, $\mathrm{err}(t,U,\Delta t, \Delta x)$, as a function of $\Delta t$, is the $l^1$ convergence error in time and $\mathrm{err}(t,U,\Delta t, \Delta x) $, as a function of $\Delta x$, is the Riemann sum associated to the $L^1$ spatial error. The numerical solution $U_{\rm{num},k}^n$ is compared to a reference solution $U_{\rm{ref}}$, which should be the exact solution. We use the solution of the linear system \eqref{eq:dustywave_pertubation_system} for $U_{\rm{ref}}$.

As illustrated in Fig. \ref{fig:dustywave_errors}, as long as the time error dominates the convergence error, it scales with $\Delta t$ (upper panel), and as long as the spatial error dominates, it scales with $\Delta x^2$ (lower panel). This demonstrates the second-order in space and first-order in time accuracy of the combined scheme.

\subsection{Disk settling} \label{sec:settling}

\begin{figure}
    \resizebox{\hsize}{!}{\includegraphics{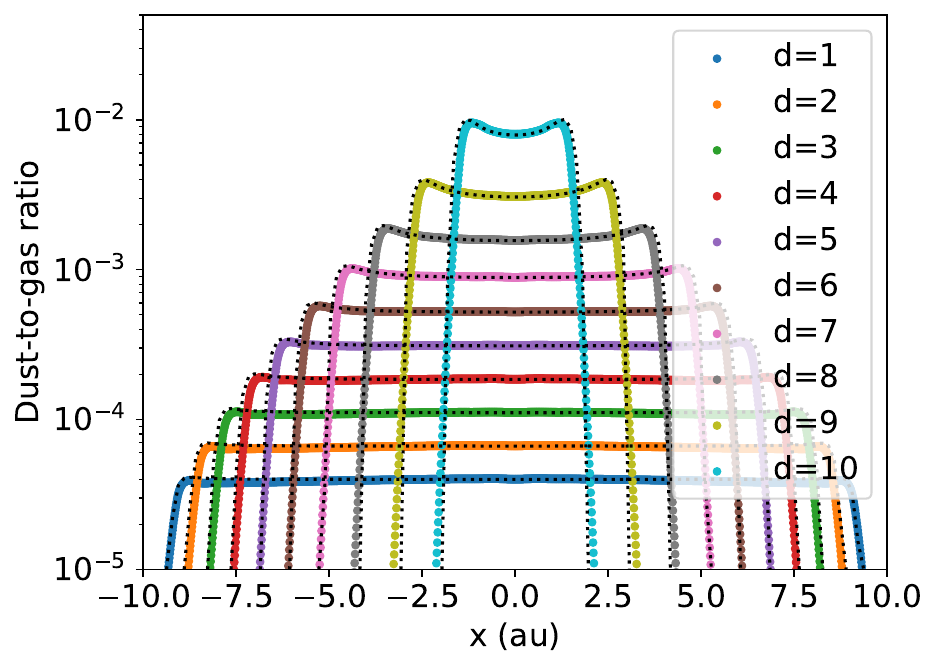}}
    \caption{Settling test after ten orbital periods, performed with the HLLgd solver at level 11 of mesh refinement (2048 cells). The time step provided by the CFL condition is divided by 5 to reach convergence ($C_{\rm{CFL}}=0.2$). Colored dots are the numerical solutions for each dust fluid, whereas the small dots are the analytical solution.}
\label{fig:settling_mrn}
\end{figure}

We assume a disk of gas at isothermal hydrostatic equilibrium with an analytical gravity acceleration. Dust grains can decouple from the gas and thus settle towards the mid-plane depending on their Stokes number, with the stopping time given by Eq. \eqref{eq:eipstein_stopping_time}. We assume an infinitely thin radial disk slice and thus the test is in 1D. We sample a size distribution with ten dust species, ranging from 100 nm ($d=1$) to 1 mm ($d=10$). This is the same setup as \citet{2018MNRAS.476.2186H} and \citet{2019A&A...626A..96L}, where the terminal velocity approach was used, as well as \citet{Lebreuilly2023}, where a multifluid approach was used. We also used the reference solution from \citet{2018MNRAS.476.2186H}. The HLLgd solver provides a satisfying solution for the whole range of dynamical coupling regimes (Fig. \ref{fig:settling_mrn}).

\section{Protostellar collapses} \label{sec:protostellar_collapses}

The conversion of dust grains to small planetesimals could happen before the class II disk is formed \citep{2018A&A...618L...3M}. Consequently, we should not exclude early planet formation scenarios, and we should explore the most embedded phases, from the protostellar collapse to class 0/I disks with their envelopes. By performing numerical simulations of a protostellar collapse, we are able to provide self-consistent initial conditions for young disks. However, some ingredients remain unknown, such as the initial dust distribution and the initial level of turbulence. The role of these parameters are highly nonlinear and, thus, we vary these two parameters in simulations. In particular, we consider which grains remain coupled to the gas and where. Indeed, for a given grain size, the degree of coupling with the gas varies depending on the local density. The stopping time is shorter in high gas density regions. The multifluid implementation allows us to treat these changes of dynamical regimes self-consistently for any dust population. Ultimately, we want to explore how grains respond to turbulence, since it is a key parameter in star formation and disk formation. Thus, we investigate the effect of turbulence on the dust enrichment conditions when forming the first hydrostatic core.

We model the initial dense core as a sphere of uniform density in solid body rotation. The gas is non-ideally coupled to the magnetic field. Here, dust represents $\theta_{d,0}=1 \%$ of the initial mass of the gas. All dust mass is represented by only one fluid characterized by one grain size. The details of the numerical setup can be found in Appendix \ref{sec:collapse_setup}. 

The degree of coupling of a grain with the gas can be parametrized by the Stokes number, which we can define in the collapse conditions as the ratio of the stopping time of the grain to the free-fall time. Following our numerical setup, it can be evaluated as
\begin{equation}
    \mathrm{St}_{d,\rm{ff}} \approx \frac{s_{\rm{grain,}d}}{1 \ \rm{mm}} \left( \frac{\rho_g}{\rho_0} \right)^{-1/2},
    \label{eq:Stokes_number_freefall_time}
\end{equation}
where $\rho_0 \approx 2.5 \times 10^{-18} ~\rm{g.cm^{-3}} $ is the density of the initial core.

We only vary the grain size of the dust fluid and the initial Mach number of the turbulence. We also test for different numerical solvers.

We split our study as follows. In Sects. \ref{sec:collapse_small} and \ref{sec:collapse_large}, we compare the multifluid implementation with the terminal velocity approach, respectively for small and large grains. In Sect. \ref{sec:collapse_turb}, we investigate the effect of the initial turbulence of the dense core and the grain size on the dynamics of the dust, and thus the dust enrichment within the first hydrostatic core and the envelope.

\subsection{Submicron grain dynamics} \label{sec:collapse_small}

\begin{figure*}
    \centering
    \includegraphics[height=0.30\textheight]{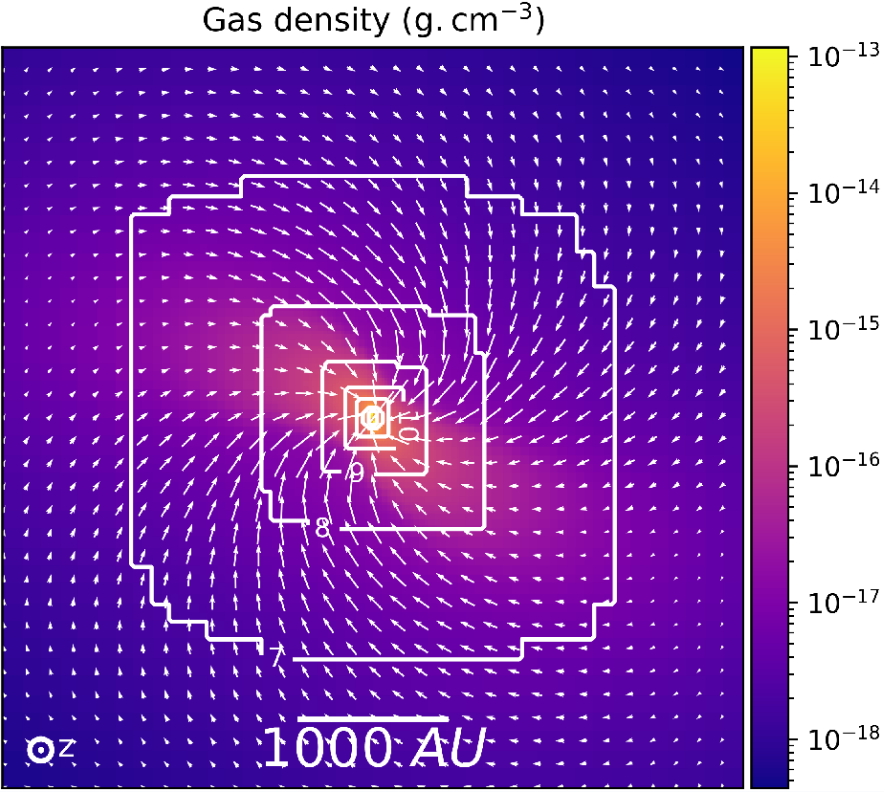}
    \includegraphics[height=0.30\textheight]{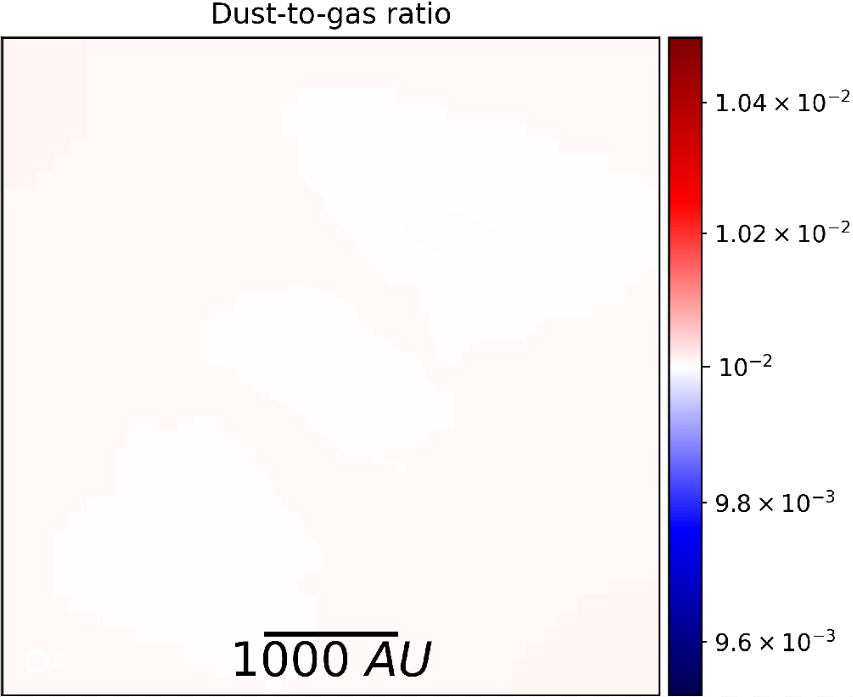}
    \caption{Collapse simulations with 100 nm grains using the terminal velocity at the formation of the first hydrostatic core ($\sim 60$kyr). The HLLD solver \citep{2005JCoPh.208..315M} is used for the (gas-dust) monofluid. Gas density is on the left (slice), which is a reference for other simulation runs as long as the feedback of the dust remains weak. Mesh-refinement levels are displayed with the contours. White arrows indicate the gas velocity. Corresponding dust-to-gas ratio map on the right, as predicted by the terminal velocity implementation.}
    \label{fig:small_grains_tva}
\end{figure*}

\begin{figure*}
    \centering
    \includegraphics[width=\textwidth]{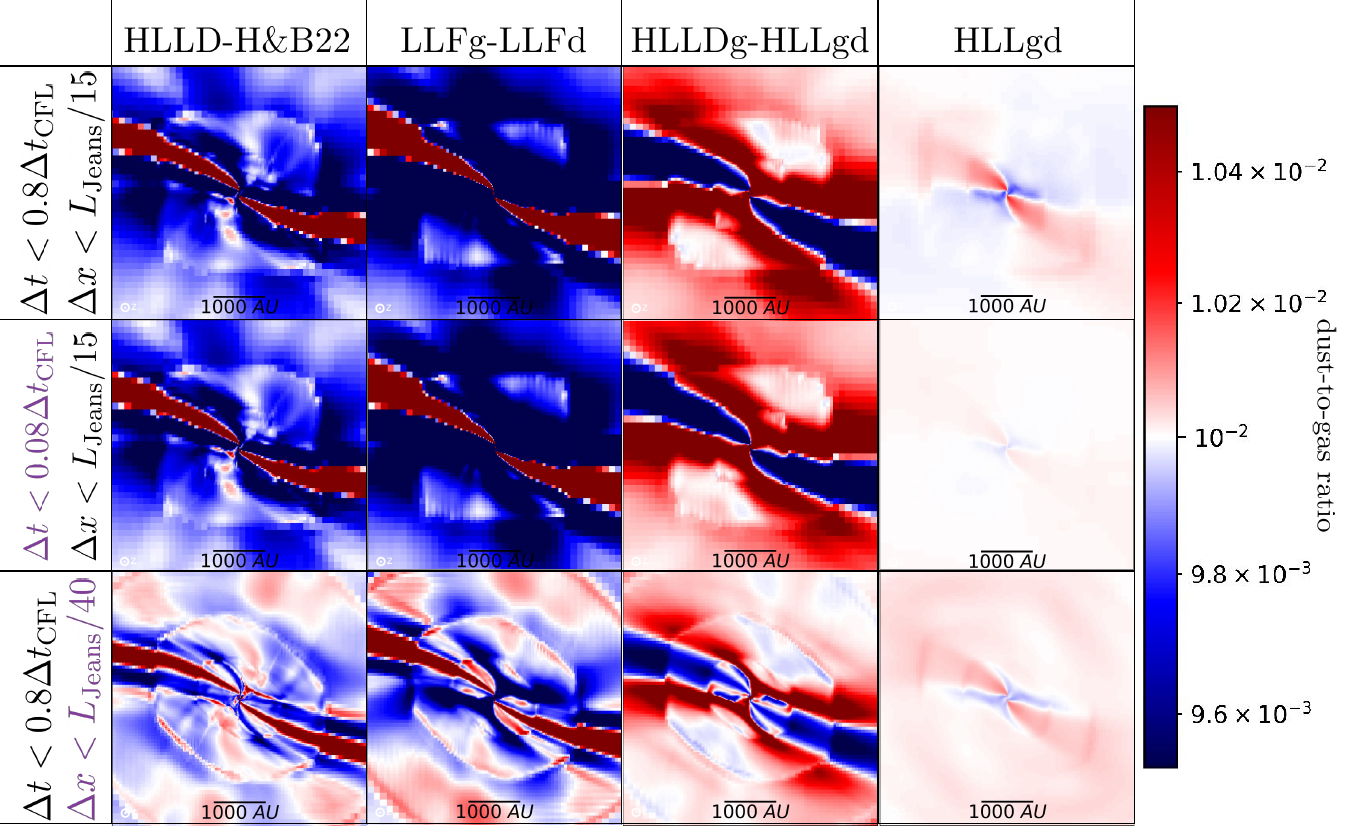}
    \caption{Collapse simulations with the multifluid implementation for different Riemann solvers (columns) and for different spatial and time resolution (lines), expressed by the safety factor of the CFL condition and the number of Jeans length per cells for the mesh refinement. HLLD is from \citet{2005JCoPh.208..315M}, H\&B22 stands for the Riemann solver from \citet{2022ApJS..262...11H}, HLLDg-HLLgd means that we use the HLLD solver for the gas and the HLLgd solver only for the dust multifluid. We choose the same colorbar scale and box size as Fig. \ref{fig:small_grains_tva} in order to compare to the terminal velocity approximation. However, for the three first solvers, some regions are saturated: for the low-resolution runs, the dust-to-gas ratio (divided by $ 10^{-2}$) in these regions vary from 0.9 to 1.2 for HLLD-H\&B22, from 0.75 to 1.3 for the LLFg-LLFd, and from 0.9 to 1.5 for HLLDg-HLLgd.
}
    \label{fig:small_grains_dust_to_gas}
\end{figure*}

\begin{figure}
    \resizebox{\hsize}{!}{\includegraphics{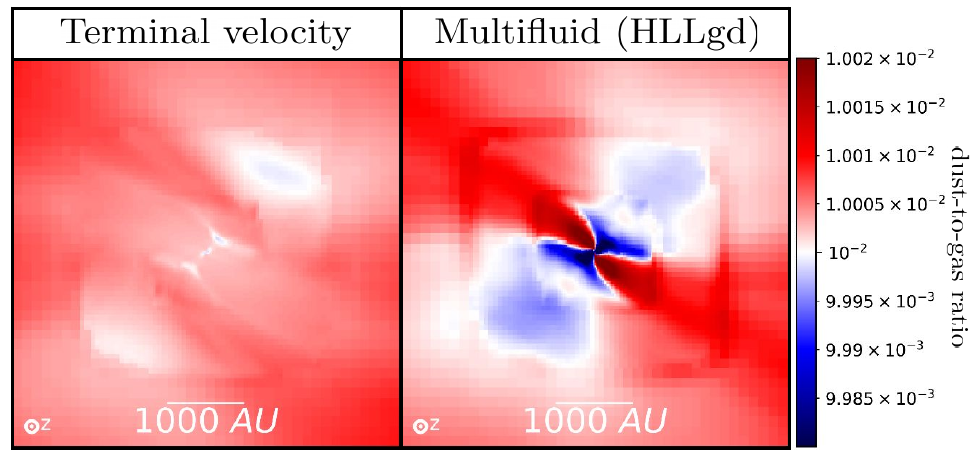}}
    \caption{Direct comparison of the dust-to-gas ratio maps between the terminal velocity simulation (Fig. \ref{fig:small_grains_tva}, right) and the multifluid simulation using the HLLgd solver with the CFL condition $\Delta t < 0.08 t_{\rm{CFL}}$ (Fig. \ref{fig:small_grains_dust_to_gas}, fourth column, second line). The colorbar scale is adapted from previous figures to emphasize the very small deviations to the initial dust-to-gas ratio of $\theta_d= 1 \%$. 
}
    \label{fig:small_grains_dust_to_gas_tva_vs_hllgd}
\end{figure}

When the stopping time is small compared to the dynamical time (which can be estimated in collapsing regions by the free-fall time), dust grains adapt their velocity to the gas dynamics. Consequently, when the gas and the dust behave as a single tightly coupled mixture, the terminal velocity approximation provides a first-order approximation of the velocity drift as a function of the stopping time and the difference in the accelerations between the gas and the dust induced by the force balance \citep{2005ApJ...620..459Y,2014MNRAS.440.2136L}. From Eqs. \eqref{eq:dust_density_conservation}-\eqref{eq:eipstein_stopping_time} with one dust species $d$ and by adding the Lorentz force $\mathbf{J} \times \mathbf{B}$ to the gas dynamics (MHD setup), the terminal velocity can be derived as

\begin{equation}
    \mathbf{V}_d - \mathbf{V}_g = t_{s,d} \frac{\nabla P_g- \mathbf{J} \times \mathbf{B}}{\rho_g+\rho_d}.
    \label{eq:terminal_velocity}
\end{equation}

This approach allows us to reduce the number of equations we have to solve. Indeed, the continuity equation becomes sufficient to fully account for the dust dynamics. The implementation of this simplified treatment of the dust dynamics in RAMSES was presented by \citet{2019A&A...626A..96L}. It was later used to model the dynamics of dust grains in nonturbulent protostellar collapses for multiple dust species in \citet{2020A&A...641A.112L}. They found that grains decouple from the gas for sizes larger than a few 100 microns.

In this section, we compare the terminal velocity approximation with the multifluid implementation. In a smooth flow and for 100 nm grains (typical size to probe the dynamics of the MRN distribution, corresponding to a Stokes number of $10^{-4}$ according to Eq. \eqref{eq:Stokes_number_freefall_time}), the terminal velocity approximation provides a reference solution presented in Fig. \ref{fig:small_grains_tva}. The dust remains very well coupled to the gas and dust-to-gas variations are very weak (less than one percent), as underlined by the choice of the colorbar scale. In Fig. \ref{fig:small_grains_dust_to_gas}, we present the results from the multifluid with different Riemann solvers.

When using the HLLD-H\&B22 solver \citep{2005JCoPh.208..315M,2022ApJS..262...11H}, the dust-to-gas ratio profile is quite different from the one obtained using the terminal velocity approximation. It is spread much more around the initial $1 \%$. Even when considering unphysical small grains (i.e., subnm grains, which should trace the gas perfectly), the dust-to-gas ratio profile does not change. Moreover, this does not change when increasing the time resolution, unless the spatial resolution is increased as well, as presented by the first column of Fig. \ref{fig:small_grains_dust_to_gas}. We obtain similar results when using individual local Lax-Friedrichs solvers for the gas and for the dust fluid (LLFg-LLFd, second column). This example demonstrates that this decoupling is artificial and it occurs because density fluxes are computed differently (LLFg and LLFd use independent wave speeds), even when the fluid velocities are the same. This artifact depends on the spatial resolution and we expect the total error to be dominated by the spatial error in the runs with individual Riemann solvers. Indeed, we expect the diffusion part of the local Lax-Friedrichs flux, which depends on the considered wave speed (Appendix \ref{sec:LLFd}), to vanish at high spatial resolution. The fluxes of LLFg-LLFd and LLFgd are asymptotically the same and therefore these two solvers are in agreement at high resolution. 

The problem is solved when using a common Riemann solver (HLLgd, last column). We can converge in time with the same spatial resolution and we can achieve a similar result to the prediction with the terminal velocity (Fig. \ref{fig:small_grains_tva}, with a more contrasted map in Fig. \ref{fig:small_grains_dust_to_gas_tva_vs_hllgd}). Indeed, for the HLLgd solver, the dust-to-gas ratio relative variations are of the order of $2 \%$ for $\Delta t < 0.8 \Delta t_{\rm{CFL}}$ and of the order of $0.2 \%$ for $\Delta t < 0.08 \Delta t_{\rm{CFL}}$, which becomes negligible compared to the variations of physical interest. We recover the advection of the low Stokes grains with the gas. In the context of protostellar collapse simulations, once the artificial decoupling is fixed by the novel Riemann solver, the multifluid code would benefit from higher order-in-time schemes as currently developed in the literature. We note that using the HLLD solver for the gas and the HLLgd flux as defined in Sect. \ref{sec:HLLgd} only for the dust fluid does not provide a satisfying solution at all (third column). This is probably due to the details of the HLLD wave fan. 

The resolution level required to obtain the agreement between the different Riemann solvers is difficult to estimate and to test. It depends on the performance of the Riemann solvers. One possible criterion consists in refining the mesh enough to properly follow the coupling between the dust and the gas. Typically, this could be $\Delta x < L_d \sim c_s t_{s,d} $, similarly to what was found in the context of SPH simulations of the dustywave \citep{2012MNRAS.420.2345L} and where $L_d$ corresponds to the recoupling length after a shock \citep{2019MNRAS.488.5290L}. In the context of a protostellar collapse, this length scales as $L_d \propto  s_{\rm{grain,}d}/ \rho $, which can be more stringent than the refinement criterion on the Jeans length $L_J \sim c_s/\sqrt{G \rho}$. For 100 nm grains, this criterion is not satisfied within the whole collapse region, whereas for 1 mm grains (next section), this mainly concerns the first hydrostatic core and the forming disk (Fig. \ref{fig:riemann_solvers_mm}). This observation depends on the mesh refinement (numerical setup in Appendix \ref{sec:collapse_setup}).

\subsection{Millimeter grain dynamics} \label{sec:collapse_large}

\begin{figure*}
    \centering
    \includegraphics[height=0.30\textheight]{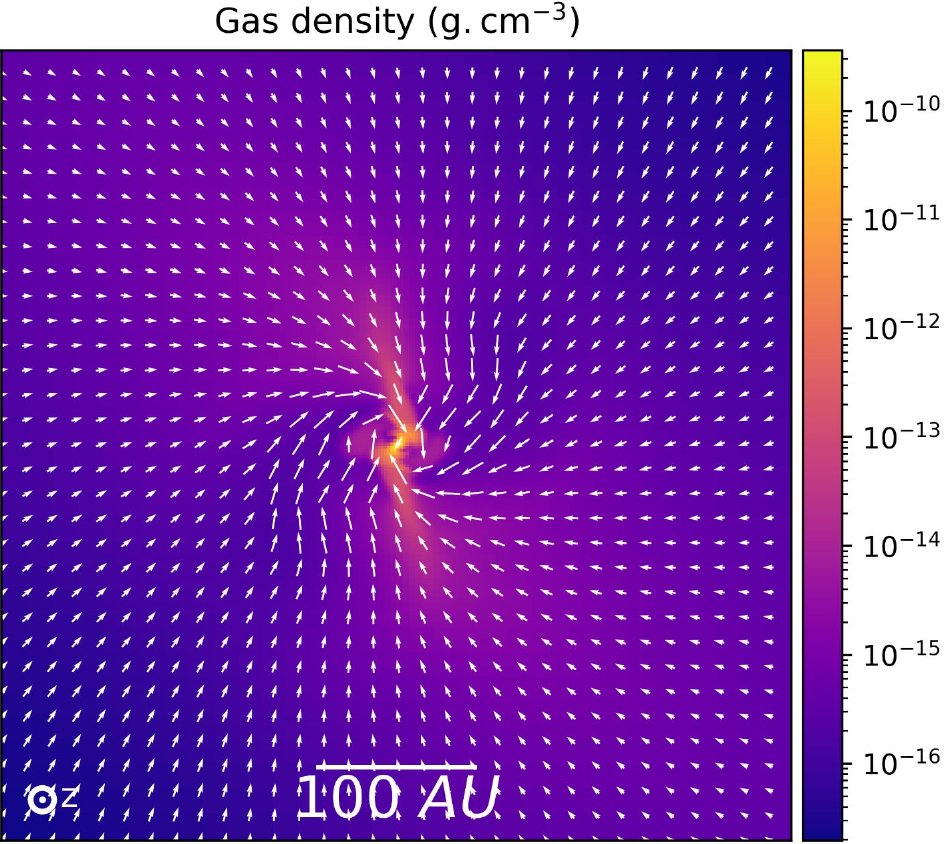}
    \includegraphics[height=0.30\textheight]{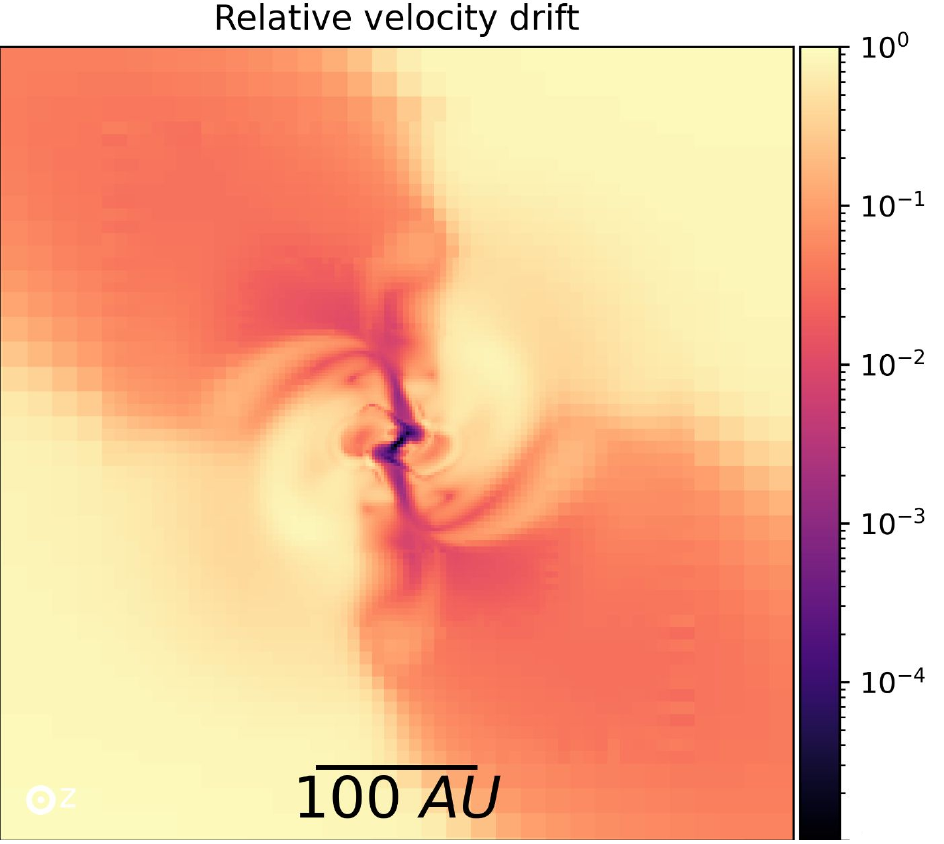}
    \caption{Multifluid simulation (HLLgd) of 1 mm grains. Gas density profile with arrows representing the gas velocities (left) and relative velocity drift $ \lVert  \mathbf{V}_d - \mathbf{V}_g \rVert / \lVert  \mathbf{V}_g \rVert$ (right).}
    \label{fig:velocity_drift_1mm}
\end{figure*}

\begin{figure}
    \resizebox{\hsize}{!}{\includegraphics{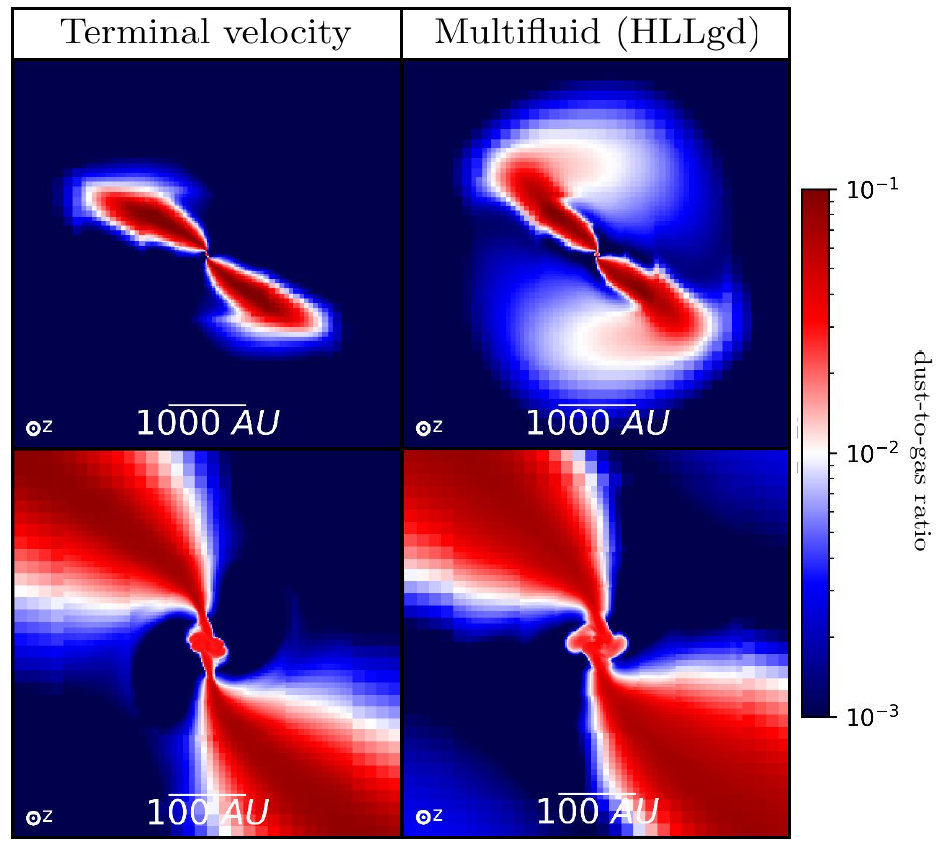}}
    \caption{Comparison at a similar epoch ($59.2$ kyr, i.e., $0.8$ kyr after the formation of the first hydrostatic core) between the terminal velocity approximation (left) and the multifluid (right). The snapshot of the multifluid is the same as in Fig. \ref{fig:velocity_drift_1mm}. Dust-to gas ratio for 1 mm grains. }
    \label{fig:Dust_to_gas_1mm_mf_vs_tva}
\end{figure}

Protostellar collapses could already host large dust grains, even though grain models remain too poorly constrained to make any firm conclusions on the size distribution. The micrometric emission and the spectroscopy of the dense regions of the molecular clouds (see \citet{2010Sci...329.1622P} for the coreshine effect and \citet{2024NatAs...8..359D} for the JWST spectroscopy) suggest that dust grains can already grow and reach micron sizes before the protostellar phase, which are above the typical sizes of the MRN distribution (5 nm to 250 nm) measured in the diffuse interstellar medium.
In the envelope of young protostars, polarized dust emission indicates the presence of grain sizes above $10~\rm{\mu m}$, found by modeling the grain alignment mechanism \citep{2019MNRAS.488.4897V}. The dust emissivity index is compatible with very large grains (up to (sub)millimeter sizes), but these conclusions are still under discussion, especially with respect to the optical properties of the grains, the dust growth mechanisms, and the dynamical origins and scales to reach such sizes \citep{2019arXiv191004652G,2023A&A...676A...4C}. Multifluid simulations accounting for the full environment (envelope, infall, core, disk, outflow) can greatly help in clarifying the latest aspect.

By selecting only 1 mm grains for protostellar collapse simulations, we explore the regime where accounting for the full inertia of dust grains matters. Indeed, the Stokes number is initially close to unity according to Eq. \eqref{eq:Stokes_number_freefall_time}, meaning that the stopping time of a 1 mm grain is comparable to the free-fall time ($\sim 42$ kyr).

As presented in Fig. \ref{fig:velocity_drift_1mm}, the velocity drift between the dust and the gas is comparable to the velocity of the gas in the lowest density regions, which breaks the terminal velocity approximation. The dust recouples to the gas in the regions of high density only (the disk and the first hydrostatic core). This is properly captured by the HLLgd solver (Appendix \ref{sec:riemann_solvers_mm}, with comparison with the H\&B22 solver). The results of the terminal velocity approximation lack of robustness and depend on the control parameters of the run (Stokes number and velocity drift limiters to avoid unrealistic dust ratios; here $ \mathrm{St}_{\rm{max}} =0.3$ and $ \lVert \mathbf{V}_d-\mathbf{V}_g \rVert <5\times 10^{5}~\rm{cm.s^{-1}}$, respectively). Therefore, the dust-to-gas ratio features are different between the two runs in the lowest density regions. More precisely, the terminal velocity approximation produces dust-to-gas ratio waves escaping the disk (Fig. \ref{fig:Dust_to_gas_1mm_mf_vs_tva}, lower panels) and a different rotating and settling envelope (Fig. \ref{fig:Dust_to_gas_1mm_mf_vs_tva}, upper panels).
Still, as captured by the terminal velocity approximation, the pressure gradients (and the Lorentz force) seem to be the main mechanism for dust enrichment. 

In these strong decoupling regions, the velocity drift is of the same order of magnitude as the sound speed. We note that correcting the stopping time, following the works of \citet{1975ApJ...198..583K} and \citet{1979ApJ...231...77D}, could partially limit such large drifts. Moreover, when using only one pressureless fluid, we cannot track a large velocity dispersion between grains, which is likely to occur for high Stokes grains in the turbulent motion of the gas. For example, the model of \citet{2007A&A...466..413O} predicts that the velocity dispersion is of the order of the gas velocity for grains whose Stokes number is close to unity.

\subsection{Dust in turbulent protostellar collapses} \label{sec:collapse_turb}

\begin{figure*}
    \centering
    \includegraphics[width=\textwidth]{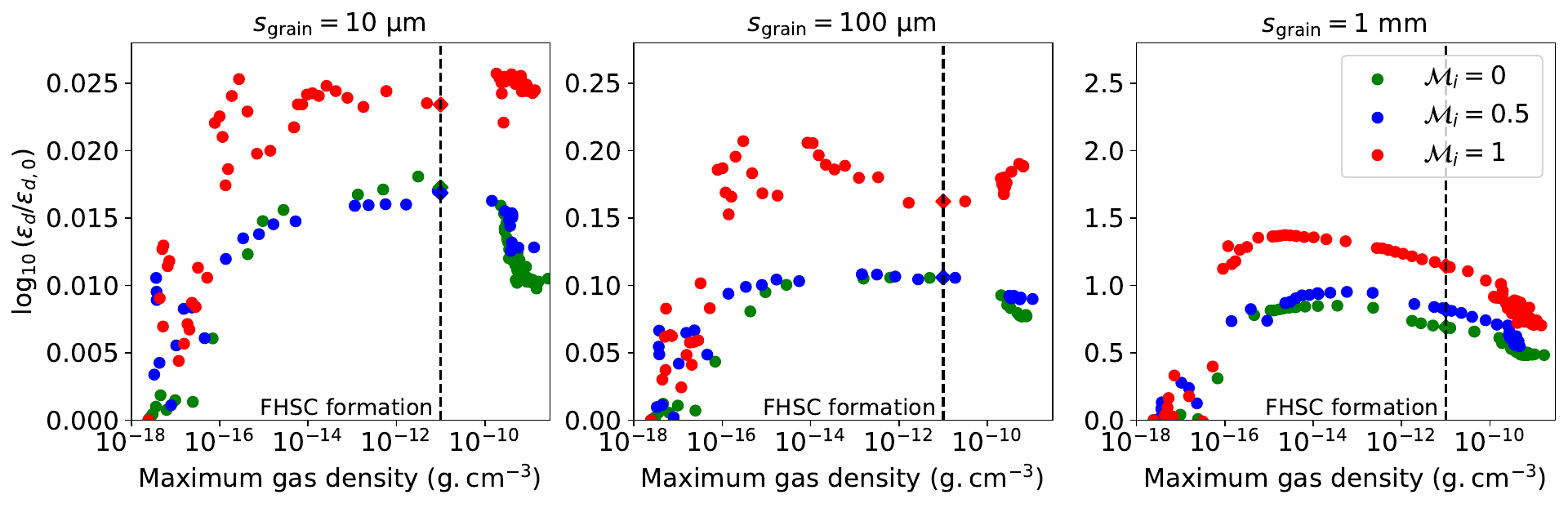}
    \caption{Evolution of the dust enrichment during the collapse. Logarithm of the normalized dust ratio at the maximum gas density, depending on the size of the grains (individual panels) and the initial turbulent Mach (green for $\mathcal{M}_i=0$, blue for $\mathcal{M}_i=0.5$, red for $\mathcal{M}_i=1$). Diamonds on the dotted line indicate the interpolated value to get the dust enrichment within the first hydrostatic core (Table \ref{tab:dust_to_gas_ratio_FHSC}).
}
    \label{fig:d2g_rhomax}
\end{figure*}

\begin{table}
    \caption{Dust-to-gas ratio within the first hydrostatic core as a function of grain size and initial turbulent Mach number.}
    \label{tab:dust_to_gas_ratio_FHSC}
    \centering
\begin{tabular}{|c|c|c|c | }
\hline
         &     $s_{\rm{grain}} = 10~\rm{\mu m}$ & $s_{\rm{grain}} = 100~\rm{\mu m}$ & $s_{\rm{grain}} = 1~\rm{mm}$ \\
\hline
       $\mathcal{M}_i=0$  & $0.01041$ & $0.0128 $ & $0.051$ \\
\hline
       $\mathcal{M}_i=0.5$  & $0.01040$ & $0.0128$ & $0.07$ \\
\hline
       $\mathcal{M}_i=1$  & $0.01056$ & $0.0146 $ & $0.16$ \\
\hline
\end{tabular}
\tablefoot{Grain sizes $s_{\rm{grain}}$ are indicated in columns and initial turbulent Mach numbers $\mathcal{M}_i$ are indicated in lines. Dust-to-gas ratios $\theta_d$ are estimated at the formation of the FHSC, that is the first time the gas density reaches $\rho_g=10^{-11}~\rm{g.cm^{-3}}$. For each run, the initial dust-to-gas ratio is $\theta_{d,0}=0.01$.
}
\end{table}

\begin{figure*}
    \centering
    \includegraphics[width=\textwidth]{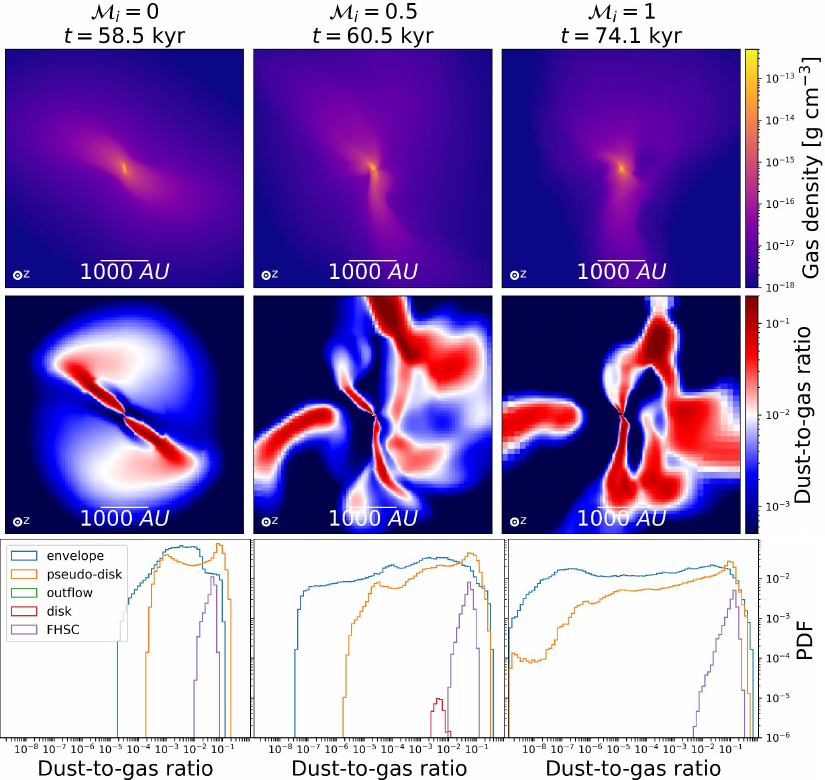}
    \caption{Collapse simulations with 1 mm grains and three initial levels of turbulence. They are presented in columns, indicated by the initial turbulent Mach number $\mathcal{M}_i$ and the snapshot time, corresponding to the formation of the first hydrostatic core ($\rho_{g, \rm{max}}=10^{-11}~\rm{g.cm^{-3}}$). In lines, the gas density map (slice), the dust-to-gas ratio map (slice), and the dust-to-gas ratio probability density function (PDF) in different regions of the collapse (FHSC, disk, outflow, pseudo-disk, envelope, if detected). The histogram is weighted by the gas mass within the cell and we use 100 log-spaced bins. This means, for example for $\mathcal{M}_i=1$, that $70 \%$ of the gas mass contains a dust enrichment lower than the initial one ($\theta_d<10^{-2}$), and about $7 \%$ of the gas mass is very dust-enriched with $\theta_d>10^{-1}$.
}
    \label{fig:turbulence_map_hist_1mm}
\end{figure*}

Measurements of molecular lines suggest that there are internal turbulent motions within protostellar dense cores which are typically subsonic \citep{1998ApJ...504..207B,2007A&A...472..519A}. Here, we investigate the impact of the initial turbulence on local dust enrichment during the protostellar collapse. In addition to the initial solid body rotation, we apply a velocity field generated in the Fourier space by random phases according the $-5/3$ power law. We varied the initial turbulent Mach $\mathcal{M}_i \in \{0,0.5,1\}$, defined by the velocity dispersion divided by the sound speed. 
We performed simulations for one dust fluid representing grain sizes of $s_{\rm{grain}} \in \{10~\rm{\mu m}, 100~\rm{\mu m},1~\rm{mm}\} $. We set the intrinsic density of the grains to $\rho_{\rm{grain}}=1 ~\rm{g.cm^{-3}}$ in this paper, but more compact grains reach typically $\rho_{\rm{grain}}=3 ~\rm{g.cm^{-3}}$. We should keep in mind that more compact grains are equivalent to larger grains because of the expression of the stopping time in Eq. \eqref{eq:eipstein_stopping_time}.

We found out that grains smaller than 10 $\rm{\mu m}$ remain highly coupled to the gas at all the tested levels of turbulence (Fig. \ref{fig:d2g_rhomax}). The 100 $\rm{\mu m}$ grains are decoupled from the gas, but the terminal velocity approximation still stands. For larger grains, the velocity drifts are such that a multifluid approach is required. This can be illustrated by modeling the evolution of the dust ratio $\epsilon_d=\rho_d/(\rho_g+\rho_d)$ as a function of the grain size (Fig. \ref{fig:d2g_rhomax}). We can see the similarity between the three panels, for each grain size. Increasing the grain size by one order of magnitude increases $\log(\epsilon_d/\epsilon_{d,0})$ by about one order of magnitude. This is compatible with the model from \citet{2020A&A...641A.112L}, which predicts that $\epsilon_d/\epsilon_{d,0} = \exp( s_{\rm{grain}}/ s_{\rm{ref}})$, where $s_{\rm{ref}}$ is a function of the density profile. Thus, $s_{\rm{ref}}$ depends on the evolution stage and the initial Mach number. The dust enrichment trend is a bit more shallow than this exponential trend. However, rough estimates from 10 $ \rm{\mu m}$ grains and 100 $\rm{\mu m}$ grains can be computed at the formation of the first hydrostatic core (FHSC) from values in Table \ref{tab:dust_to_gas_ratio_FHSC}. We found $s_{\rm{ref}} = 250-410~\rm{\mu m}$ at $\mathcal{M}_i=0$ and $0.5$ and $s_{\rm{ref}} = 180-270~\rm{\mu m} $ at $\mathcal{M}_i=1$. These sizes mean that initial turbulence promotes dust enrichment in the FHSC as if the effective size of the grains is lowered. The dust enrichment model is no longer valid for millimeter grains, most likely because most of the assumptions required to obtain this relationship, such as strong coupling and weak back-reaction, cannot be satisfied. However, the dust enrichment remains an increasing function of the size and initial turbulence. The initial turbulence delays the formation of the FHSC, thereby giving more time to the dust to fall and to settle. This is underlined in Fig. \ref{fig:d2g_rhomax}, where the dust-to-gas ratio reaches a maximum before the gas starts to form a dense hydrostatic core. Then the dust-to-gas ratio decreases as less dust-enriched material starts to fall and to mix in the inner region.

We extended our analysis to lower density regions. We computed the probability density function (PDF) of the dust-to-gas ratio in different regions of the collapse, within a radius of 2500 AU. We select cells belonging to the first hydrostatic core (FHSC), the disk, the pseudo-disk, the outflows, and the envelope following the criteria used in \citet{2020A&A...641A.112L}, with the references therein. We caution that the selection criterion for the cells belonging to the FHSC is $\rho_g>10^{-12.5}~\rm{g.cm^{-3}}$. The PDF of the dust-to-gas ratio for 10 $\rm{\mu m}$ grains and 100 $\rm{\mu m}$ grains is narrow and peaks around the enrichment within the FHSC (details in Appendix \ref{sec:turb_10um_100um}), as expected; whereas the PDF for 1 mm grains is particularly extended (Fig. \ref{fig:turbulence_map_hist_1mm}). Therefore, we dedicate the rest of the section to the description of the dust enrichment and depletion of 1 mm grains. We present in Fig. \ref{fig:turbulence_map_hist_1mm} the gas density profile (the feedback from dust grains may not be negligible) and the corresponding dust-to-gas ratio map.

Without any initial turbulence, the main mechanism for dust enrichment in low density regions seems to be the settling due to the pressure gradient created by the stable dense regions of the pseudo-disk. On the contrary, initial turbulence form dust-rich structures in the envelope, where grains are highly decoupled. These structures seem to be decorrelated from the gas density profile in the envelope at the end of the first protostellar collapse phase. Indeed, even though the resulting pseudo-disk is distorted, the gas profile is smooth (upper panels); this is most likely because thermal (and magnetic) pressure provides support against compression, which is not the case for the dust fluid. The probability density function provides the typical maximum dust-to-gas ratios in the low-density regions (envelope and pseudo-disk). We obtained $\theta_d=10\%, 25\%,40\%$ for $\mathcal{M}_i=0,0.5,1$, respectively. This very promising enrichment of large grains in some locations of the envelope and the pseudo-disk also means that most of these regions becomes highly depleted in dust.

With such enrichment in the envelope and the hydrostatic core, the feedback of the dust on the gas, via gravity and drag, cannot be neglected and it strongly affects the formation of the disk. Indeed, it modifies the balance between the rotational and the gravitational energy, from the gas and the dust, and the thermal energy, only from the gas. These simulations are possibly extreme, because of the initial dust distribution and resulting decoupling and feedback. However, it demonstrates the versatility and robustness of the current multifluid implementation regarding the vast range of dynamical scales.

In dense regions (the first hydrostatic core and the disk) or in very dust-enriched regions (potentially the envelope), the grain coagulation time becomes shorter than the dynamical time scales and, thus, it becomes a dominant mechanism for the evolution of the dust distribution. In protoplanetary disks, the dust continuum emission indicates that grains reach submillimeter sizes \citep{2015ApJ...809...78K}. We should therefore account for dust growth by using, for instance, the method of \citet{2021MNRAS.501.4298L} based on the Eulerian multifluid approach. 
This requires accurate hydrodynamical velocity drifts between the dust species. In that respect, the common Riemann solver should also help in preventing unphysical velocity drifts when advecting the velocity components. 
Coupling the hydrodynamics and the growth of the dust distribution should offer a better hint at the conditions of early planet formation.

\section{Conclusion} \label{sec:conclusion}

We implemented a multifluid method in the RAMSES code for the multiscale physics of dusty flows. We tested our implementation, including up to ten dust fluids in various coupling regimes (Sect. \ref{sec:settling} and Appendix \ref{sec:dustyjeanswave}). We emphasize the difficulty in capturing the coupling regimes between the dust and the gas, particularly during a protostellar collapse. We present a novel Riemann solver based on the HLL approach to deal with the coupling regimes consistently.

\begin{itemize}
    \item Current multifluid methods use individual Riemann solvers for each fluid with corresponding truncation errors. In the strong coupling regime, this leads to unbalanced density advection steps and, thus, unphysical dust-to-gas ratio variations.
    \item When using individual solvers, we found out that the dust enrichment within the first hydrostatic core cannot be studied properly. Indeed, with this strategy, the enrichment of MRN grains, which are well coupled to the gas, is dominated by spatial errors (Sect. \ref{sec:collapse_small}), where as for millimeter grains, they cannot properly recouple to the gas within the first hydrostatic core (Appendix \ref{sec:riemann_solvers_mm}).
    \item Instead, we use a common Riemann solver for the gas and the dust multifluid, which is based on the HLL wave fan of the gas. In the strong coupling regime, this solution bypasses known problems with Eulerian methods for pressureless fluids, such as situations of converging flows (Appendix \ref{sec:dustywave_amr}).
    \item This novel solver eliminates the spatial truncation error observed for a tightly coupled gas and dust mixture. It can reproduce the results of the terminal velocity approximation with reasonable time resolution. This solver allows us to go beyond the terminal velocity approximation and to study situations where accounting for the inertia of dust grains self-consistently is necessary; for instance, in shocks (Appendix \ref{sec:dustyshock}).
    \item Here, we use a solver switch to deal with weak coupling regimes. The switch criterion is purely based on the fluid kinematics or, more precisely, whether the dust velocities belong to the influence area of the gas, modeled by the wave fan (Sect. \ref{sec:HLLgd}). This strategy limits numerical diffusion. We have not identified any issue with this switch at this time in the context of protostellar collapses.
    \item The common Riemann solver could provide unnecessarily more numerical diffusion than individual Riemann solvers, for instance, in the case of weakly coupled grains whose velocity drift remains small. Individual Riemann solvers could be a preferential solution in situations of weak coupling (without changes of regimes), at high spatial resolution compared to the coupling scales, or when the dust-to-gas ratio is not a quantity of interest (compared to the precision on the gas dynamics with HLLD for instance). Even though these conditions are not met in protostellar collapses, this needs to be investigated in future work. In particular, the dust enrichment we found in some regions of the collapse could favor the development of dynamical instabilities, such as the streaming instability \citep{2005ApJ...620..459Y} and the resonant drag instabilities \citep{2018MNRAS.477.5011S}. We highlight the fact that small-scale dust enrichment, turbulence, and dynamical instabilities are not resolved in the simulations we present in this paper. Investigating small-scale dynamics using the conditions set by the multi-scale collapse simulations is a direct perspective of the code. Moreover, it would allow us to test the limits of the solvers in another context.
\end{itemize}

In this paper, we emphasize the importance of accounting for the physics of a dust and gas mixture to model the hydrodynamical solver, even when using a fractional step method as we do. This is of key importance to limit errors on the dust-to-gas ratio, a central parameter for triggering dynamical instabilities such as the streaming instability and the resonant drag instabilities, which are potential paths to planetesimal formation. While current works are mainly focused on developing modular and sophisticated high-order-in-time schemes, these multifluid implementations could suffer from truncation errors on the dust-to-gas ratio dominated by the spatial resolution.

Our multifluid method is well-suited to simulations of turbulent protostellar collapses, which are multi-scale and multi-physics. We found that the dust enrichment within the first hydrostatic core is an increasing function of the grain size and the initial turbulence. Grains under $100~\rm{\mu m}$ remain well coupled to the gas while 100 $\rm{\mu m}$ grains are enriched within the first hydrostatic core (between $20 \% $ and $50 \%$ of dust-to-gas ratio variations). This is in agreement with the terminal velocity approximation. This underlines the role of the force balance between the dust and the gas. Millimeter grains significantly drift relative to the gas within the envelope. Therefore, modeling their inertia thanks to the multifluid approach is necessary. In the presence of turbulence in the initial core, dust can spread in the envelope and form enriched filaments. Millimeter grains also fall faster than the gas and enrich the inner region very early on, that is, prior to the formation of the first hydrostatic core and later fed by less enriched material. This dust mass back-reacts on the gas and affects disk formation as a result. This is a promising avenue for testing early planet formation scenarios.

\begin{acknowledgements}

We thank the referee for their comments, which contribute in strengthening the quality of the paper. We thank Leodasce Sewanou, Benoît Commerçon, Guillaume Laibe, Geoffroy Lesur and Maxime Lombart for discussions on numerical methods for dust evolution. We also thank Adnan Ali Ahmad for his advice and help in post-processing the simulations. This research has received funding from the European Research Council synergy grant ECOGAL (Grant: 855130).

\end{acknowledgements}
%-------------------------------------------------------------------
\bibliographystyle{aa}
\bibliography{ref}

\begin{appendix}

\section{Local Lax-Friedrichs solver for dust fluids} \label{sec:LLFd}

We define hereafter the local Lax-Friedrichs solver for individual dust fluids, denoted as LLFd. The local Lax-Friedrichs solver (Rusanov flux) was already presented and implemented in the context of pressureless fluids for example in \citet{2024ApJS..271....7K}. We consider for the Riemann problem, the left and right states of the dust fluid $U_{d,L}$ and $U_{d,R}$. The normal components of the dust velocity are $V_{d,L}$ and $V_{d,R}$. and the fluxes are denoted by $F_{d,L}= F_d(U_{d,L})$ and $F_{d,R}= F_d(U_{d,R})$. The fluxes for each dust species $d \in \{1,\mathcal{N} \}$ are formally defined as

\begin{equation}
    F_{LLF,d} := \frac{1}{2} ( F_{d,L} + F_{d,R}  ) - \frac{S_{LLF,d}}{2} ( U_{d,R} - U_{d,L}  ).
\end{equation} 

The term involving $S_{LLF,d}$ corresponds to the diffusion part of the solver. $S_{LLF,d}$ should be an estimate of the maximum wave speed in the fluid (see discussion in Sect. \ref{sec:HLLgd}). The answer is not straight-forward since the homogeneous system is not hyperbolic. Considering an isothermal pressure in the dust fluid with corresponding sound speed $c_d$ leads to a hyperbolic system characterized by its eigenvalues $\{V_d-c_d, V_d,+c_d\}$. Therefore, considering the limit $c_d \to 0$, we should use the normal velocity $V_d$ as the typical wave speed. We define the speed associated to the local Lax-Friedrichs solver for each dust fluid as

\begin{equation}
    S_{LLF,d}= \max{(|V_{d,L}|, |V_{d,R}|)}.
\end{equation}

\section{Dustybox: Time sequence of the drag scheme} \label{sec:dustybox_sequence}

\begin{figure}
    \resizebox{\hsize}{!}{\includegraphics{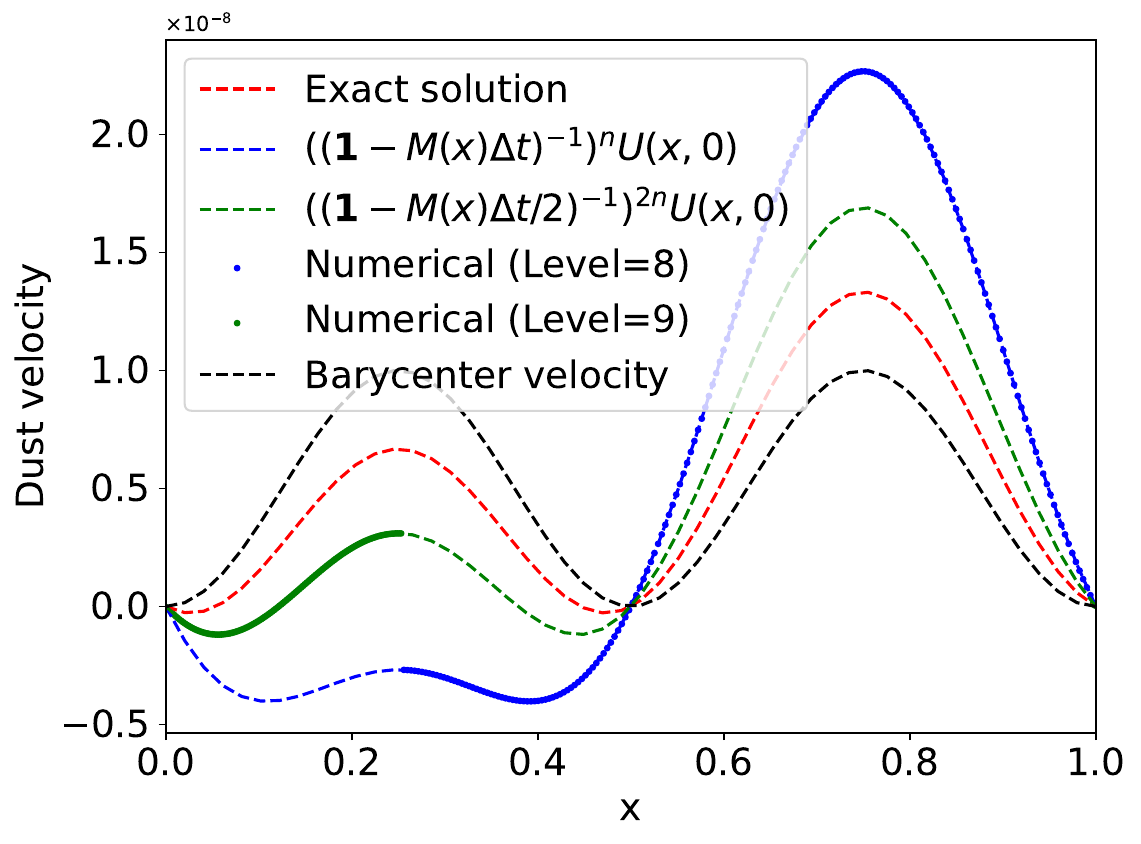}}
    \caption{Numerical solution of the dust velocity within the box domain, where points belonging the level 8 of mesh refinement are in blue, and points belonging to the level 9 are in green. The final time $t= 0.103$, corresponds to $n=33$ steps at the level 8 of mesh refinement.}
    \label{fig:dustybox_box}
\end{figure}

In this section, we derive the sequence produced by two drag schemes at each time step. The exact drag scheme, following Eq. \eqref{eq:drag_exponential}, provides, over one time step $\Delta t=t_{n+1}-t_n$,
\begin{equation}
    U(x,t_{n+1})=\exp(M(x)\Delta t) U(x,t_n).
\end{equation}
Obviously, the sequence leads to the exact solution whatever the choice of the time step
\begin{equation}
    U(x,n \Delta t)=(\exp(M(x)\Delta t))^n U(x,t_n)=\exp(M(x) n\Delta t) U(x,0).
    \label{eq:exponential_sequence}
\end{equation}

On the other hand, the first-order implicit Euler scheme is defined over $\Delta t$ as
\begin{equation}
    U_{1}(x,t_{n+1})=U_1 (x,t_{n}) + M(x) U_{1}(x,t_{n+1}) \Delta t,
\end{equation}
and thus the scheme is formally
\begin{equation}
    U_{1}(x,t_{n+1})=(\mathbf{1} - M(x) \Delta t)^{-1} U_1 (x,t_{n}).
\end{equation}
Consequently, it produces the sequence
\begin{equation}
    U_1(x,n \Delta t)=((\mathbf{1} - M(x) \Delta t)^{-1})^n U (x,0).
\end{equation}

In mesh-refined levels, we can decide to subcycle in time. When doing so, a time step $\Delta t$ is split into 2 steps of $\Delta t/2$. Thus, for these points, the first-order implicit Euler scheme over one time step $\Delta t$ produces
\begin{equation}
    U_{2}(x,t_{n+1})=(\mathbf{1} - M(x) \Delta t/2)^{-1} \times (\mathbf{1} - M(x) \Delta t/2)^{-1} U_2 (x,t_{n}),
\end{equation}
and thus the sequence
\begin{equation}
    U_{2}(x,n \Delta t)=((\mathbf{1} - M(x) \Delta t/2)^{-1})^{2n} U (x,0).
\end{equation}

We can compute these matrices and sequences, and compare to the result of the implemented drag scheme. The setup is the one presented in Sect. \ref{sec:dustybox} and here we explicitly present the solution within the box domain with two levels of mesh refinement (level 8 in green and level 9 in blue in Fig. \ref{fig:dustybox_box}). The final time $t = 0.103$ corresponds to $n = 33$ steps at the level 8 of
mesh refinement and $2n=66$ steps at the level 9, thanks to time subcycling. This is the CFL condition at the level 8 of mesh refinement if the hydrodynamics was activated, with $C_{\rm{CFL}}=0.8$ and $C_s=1$ (see the dustywave test, Sect. \ref{sec:dustywave}).

In Fig. \ref{fig:dustybox_box}, we can see the impact of the small local density perturbation on the velocity profile, because the final time is longer than the local damping time, that is when $K_d(1/\rho_g(x) + 1/\rho_d(x))t>1$. We recover the result of the first-order implicit Euler scheme with $((\mathbf{1} - M(x) \Delta t)^{-1})^n U(x,0)$ for the coarse level and $((\mathbf{1} - M(x) \Delta t/2)^{-1})^{2n} U(x,0)$ for the subcycled level. Consequently, a discontinuity appears between subcycled levels, unless the exact solver (exponential operator) is used (Eq. \eqref{eq:exponential_sequence}).

This test is stronger than a time convergence test since the expected truncation error is also tested.

\section{Dustywave appendices}

\subsection{Spatial error when coupling a finite-volume method with an ODE solver for the drag step} \label{app:FV_drag}

A finite-volume method lies on the value of conserved quantities, averaged within a cell. Thus, we illustrate by volume-averaging the drag step in Eq. \eqref{eq:dust_momentum_conservation}, and by using the form of the drag force in Eq. \eqref{eq:drag_force_dustywave}.

\begin{equation}
    \langle \rho_d v_d \rangle_I (t_2) = \langle \rho_d v_d \rangle_I (t_1) + \int_{t_1}^{t_2} \langle K_d (v_g - v_d) \rangle_I (t) dt
\end{equation}

If $K_d$ is constant, the source term that needs to be time-integrated is
\begin{equation}
\begin{aligned}
    \langle F_{g \to d} \rangle_I &= K_d  \langle (v_g - v_d) \rangle_I = K_d \left( \frac{\langle \rho_g v_g \rangle_I}{ \langle \rho_g \rangle_I } - \frac{\langle \rho_d v_d \rangle_I}{ \langle \rho_d \rangle_I }  \right) \\
    &+ K_d \left( \frac{1}{\rho_d} \frac{\partial \rho_d  }{\partial x}\frac{\partial v_d  }{\partial x}(0)  -  \frac{1}{\rho_g} \frac{\partial \rho_g  }{\partial x}\frac{\partial v_g  }{\partial x}(0) \right) \frac{\Delta x^2 }{12}  +o(\Delta x^2).
\end{aligned}
\end{equation}

The latter expression shows that estimating the velocity fields by diving the momentum by the density contributes to a second-order spatial error on the drag term. Another common source of spatial error is made when the volume-averaged value is considered to be the value at the center of the cell (initialization of the fields, comparison with a reference solution as done in Eq. \eqref{eq:convergence_error}), which contributes to a second-order spatial error according to the midpoint rule.

 \subsection{Mesh refinement and time subcycling} \label{sec:dustywave_amr}

\begin{figure*}
    \centering
    \includegraphics[width=\textwidth]{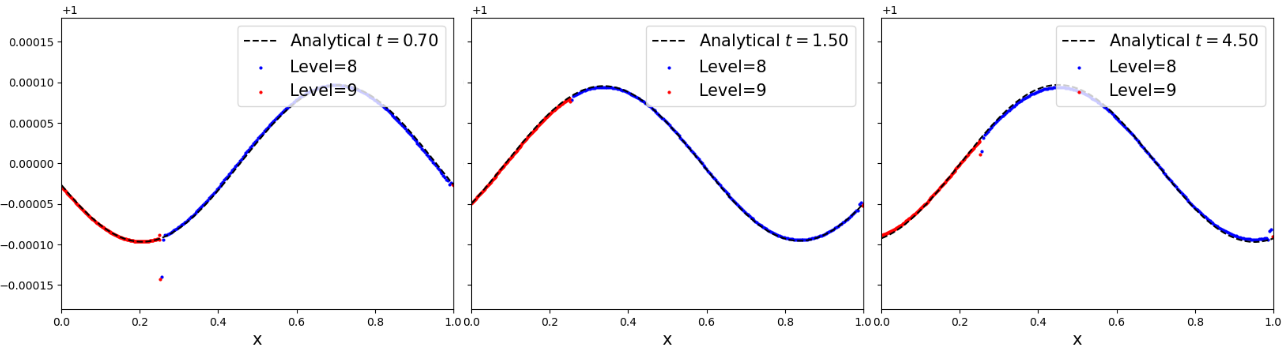}
    \caption{Dustywave test with two subcycled levels of mesh refinement. Dust density at $t=0.7,1.5,4.5$. The Riemann solvers are HLL for the gas and LLF for the dust (HLLg-LLFd).}
\label{fig:dustywave_evolution_MR_subcycling}
\end{figure*}

\begin{figure}
    \resizebox{\hsize}{!}{\includegraphics{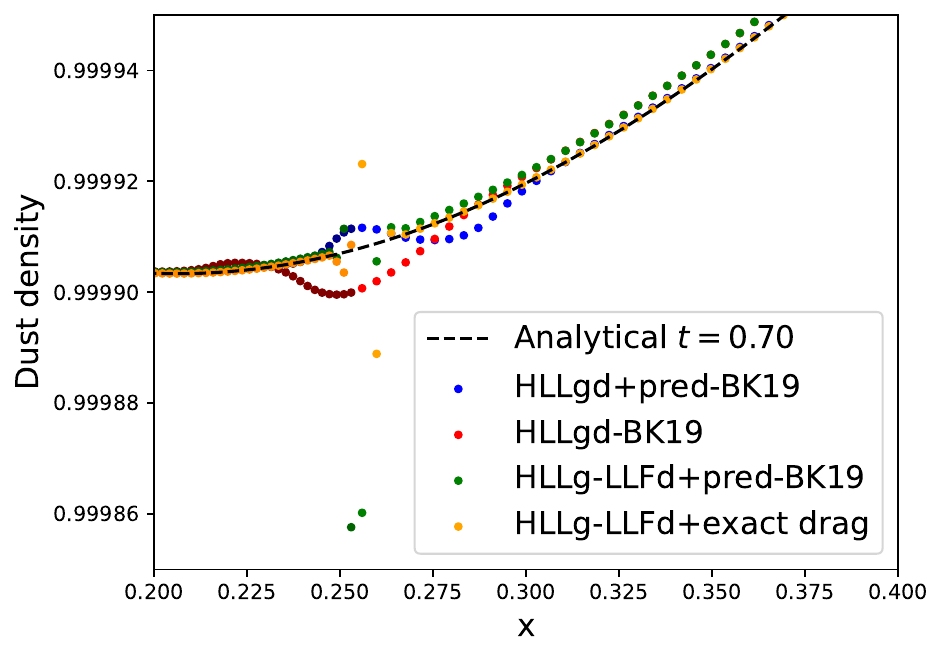}}
    \caption{Zoom-in on the dust density at $t=0.7$ for different solvers, represented by different colors (with a darker shade for the level 9 of mesh refinement). In the legend, BK19 refers to the drag solver from \citet{2019ApJS..241...25B} and pred means that we include the drag in the predictor step of the hydro solver.}
\label{fig:dustywave_solvers}
\end{figure}

In Fig. \ref{fig:dustywave_evolution_MR_subcycling}, we present three snapshots of the dustywave test with two subcycled levels of refinement. A glitch on the dust density appears in the cells in the neighborhood of the level change. As shown by the different snapshots, it oscillates on a short time scale as the dustywave propagates. By adding an ad hoc pressure to the dust fluid and by vanishing the pressure in the gas, we found that this behavior is not a bug, but a limit of the scheme to deal with pressureless fluids. We emphasize that, in this linear test, the dust density has no significant feedback on the system. Thus, the dust density is advected according to the velocity field. 
 Velocity changes can generate density peaks but pressure support could help in limiting the formation of these peaks. In the case of the pressureless dust, these small scales density structures can persist and propagate at its group velocity in the domain, which could make them difficult to track. Even when increasing the spatial resolution, this glitch is present, similarly to the case of Gibbs phenomena. When such errors pass the level boundary, an interpolation error is made. This glitch is easily seen when activating the time subcycling, but it cannot be solved by the exact drag solver (Fig. \ref{fig:dustywave_solvers}, in orange). This underlines the fact that velocity discontinuities at the interface, as presented in the dustybox test, are not the main mechanism that generates this glitch. Indeed, the same discontinuity is present in the gas velocity. 
 
 We observe a linear decrease of the glitch with the time step. Moreover, using the common Riemann solver HLLgd, whose wave fan is here mainly given by the gas sound speed, softens the glitch but at the cost of numerical diffusion (Fig. \ref{fig:dustywave_solvers}, in red). However, thanks to this numerical viscosity, improving the precision of the drag solver helps in reaching a satisfying solution (Fig. \ref{fig:dustywave_solvers}, in blue).

 We face a similar situation in the disk settling test (Sect. \ref{sec:settling}). A shock feature appears at the mid-plane of the disk when using an individual Riemann solver for the dust, which is particularly visible at low resolution. This is because the mid-plane separate two regions with the velocity change of sign $v_{d,x}(x<0)>0$ and $v_{d,x}(x>0)<0$. Dust should not form a shock in this high density region, because it is very well coupled to the static gas (the terminal velocity approximation stands), but an individual Riemann solver treats it as a problem of converging dust flows at the interface.

\section{Dustyjeanswave} \label{sec:dustyjeanswave}

\begin{table}
    \caption{Initial conditions of the dustyjeanswave.}
    \label{tab:multijeanswave_initial_conditions}
    \centering
\begin{tabular}{c|c}
        Parameter &     Value\\
\hline
         $(\delta \rho_g)_0$ & $2.7314798795353203\times 10^{-5}$ \\
         $\phi_g$ & $0.05058203158724116$ \\
         $(\delta \rho_1)_0$ & $1.7343406399117649\times 10^{-6}$ \\
         $(\delta \rho_2)_0$ & $1.689443433957737\times 10^{-6}$ \\
         $(\delta \rho_3)_0$ & $2.2962419848802065\times 10^{-6}$ \\
         $(\delta \rho_4)_0$ & $1.189737412609774\times 10^{-5}$ \\
         $(\delta \rho_5)_0$ & $1.3702893874930756\times 10^{-5}$ \\
         $\phi_1$ & $-2.6633323209942565$ \\
         $\phi_2$ & $-2.784967044742389$ \\
         $\phi_3$ & $2.4001985938943213$ \\
         $\phi_4$ & $0.7389309844855019$ \\
         $\phi_5$ & $0.12295937958860421$ \\
         $(v_g)_0$ & $5.616883358274461\times 10^{-5}$ \\
         $\psi_g$ & $-0.0723773480013628 $ \\
         $(v_1)_0$ & $7.132828728400922\times 10^{-6}$ \\
         $(v_2)_0$ & $6.948179834704889\times 10^{-6}$ \\
         $(v_3)_0$ & $9.443762326844544\times 10^{-6}$ \\
         $(v_4)_0$ & $4.893037158114604\times 10^{-5}$ \\
         $(v_5)_0$ & $5.6355938876176875\times 10^{-5}$ \\
         $\psi_1$ & $-2.7862917005828582$ \\
         $\psi_2$ & $-2.907926424331034$ \\
         $\psi_3$ & $2.2772392143057014$ \\
         $\psi_4$ & $0.6159716048968983$ \\
         $\psi_5$ & $0.0$ 
\end{tabular}
\tablefoot{These parameter values describe the eigenmode corresponding to the eigenvalue $-1.584688946477482-12.82288981648163i$ of matrix \eqref{eq:dustyjeanswave_matrix}.}
\end{table}

\begin{figure}
    \resizebox{\hsize}{!}{\includegraphics{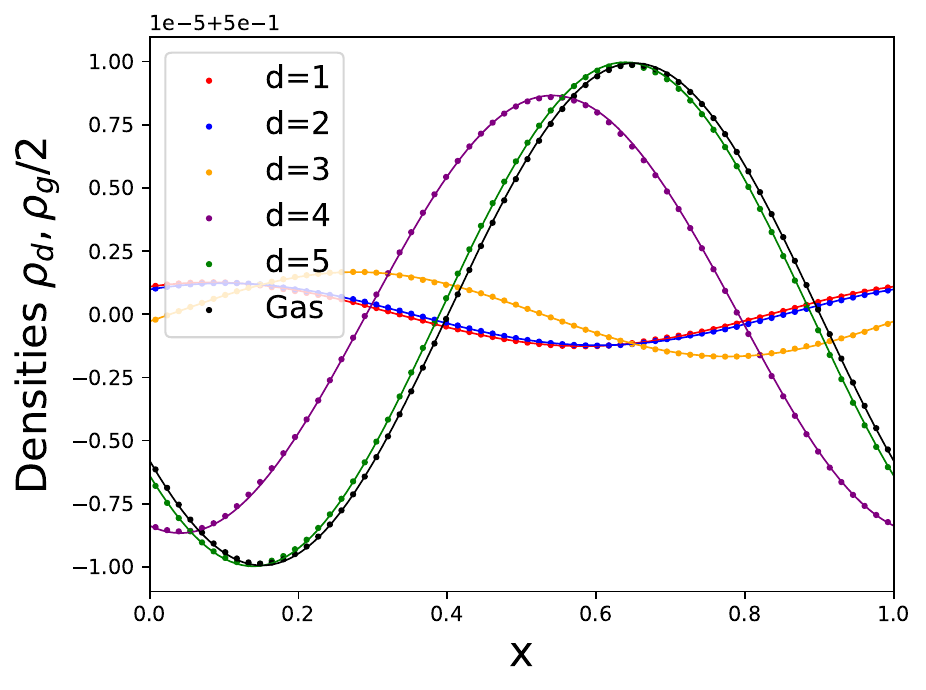}}
    \caption{Densities at $t=0.2$ (which corresponds to $ \sim 0.5$ crossing time of the wave) of the five dust species and of the gas (divided by 2) along with the eigenmode described in Table \ref{tab:multijeanswave_initial_conditions}. Dots are the numerical solutions. Continuous lines are the linear solutions. The time step is divided by a factor of 10 ($C_{\rm{CFL}}=0.1$) to achieve the convergence in time for this resolution (level 6, 64 points).
    }
    \label{fig:multijeanswave}
\end{figure}

In this test, several dust fluids and the gas interact via gravity and drag, using the completeness of the equations Eqs. \eqref{eq:dust_density_conservation}-\eqref{eq:gas_energy_conservation}. For one dust species, the evolution of the state $W=(\delta \rho_g, \delta \rho_d, v_g, v_d)^T$ is given by the system $d_t W  = M_J W$ with

\begin{equation}
    M_J=
    \begin{pmatrix}
        0 & 0 & -ik \rho_g & 0 \\
        0 & 0 & 0 & -ik \rho_d \\
        -ik C_s^2/\rho_g - 4\pi G /(ik)  & -4\pi G/(ik) & -K_d/\rho_g & K_d/\rho_g \\
        - 4\pi G /(ik)  & -4\pi G/(ik) & K_d/\rho_d & -K_d/\rho_d
    \end{pmatrix}.
    \label{eq:dustyjeanswave_matrix}
\end{equation}

Contrary to the dustywave test (Sect. \ref{sec:dustywave}) for which the density of the dust is mostly advected according to its velocity field, here the dust density can strongly feed back the dynamics of the system. We replicate the linear analysis of the Jeans instability extended to a mixture of gas and several dust fluids. In particular, we simulate one of the damping mode of the system with five dust fluids. We chose $(\rho_g)_0=1$ for the gas and $(\rho_1)_0=(\rho_2)_0=(\rho_3)_0=(\rho_4)_0=(\rho_5)_0=0.5$ for the dust fluids such that the contribution of dust to the gravity is high. We chose $K_1=0.01$, $K_2=0.1$, $K_3=1$, $K_4=10$, and $K_5=100$ for the drag coefficients of each dust species such that the test covers a vast range of Stokes numbers simultaneously. We chose $C_s=3$ to get the propagation of a compressible wave. We set $k=2 \pi$ and $G=1$.

We selected the propagating and damped eigenmode of $M_J$ (extended to five dust species) described in Table \ref{tab:multijeanswave_initial_conditions}. We successfully recovered the dynamics of the multifluid mixture, as presented in Fig. \ref{fig:multijeanswave}.

\section{Shock in a dust and gas mixture} \label{sec:dustyshock}

\begin{figure}
    \resizebox{\hsize}{!}{\includegraphics{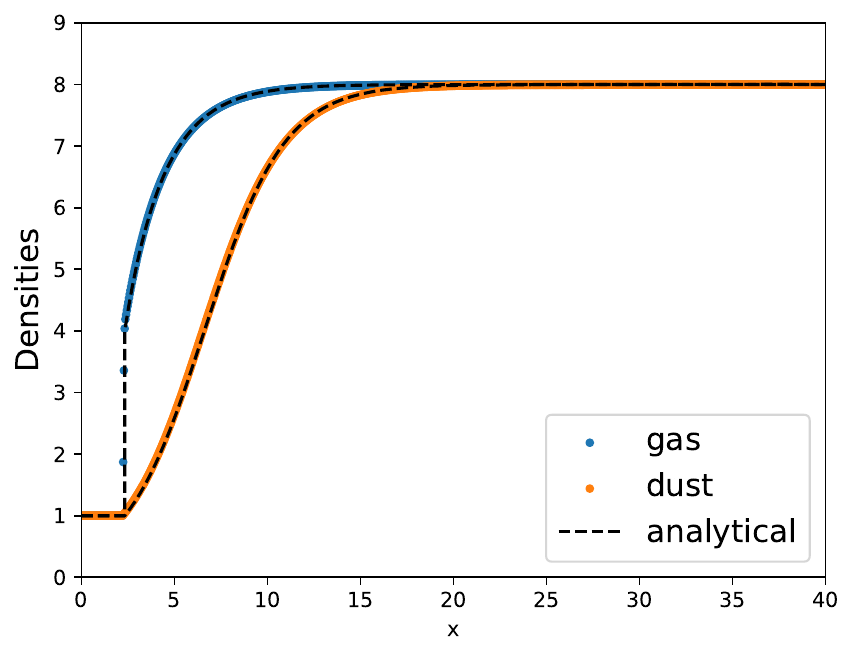}}
    \resizebox{\hsize}{!}{\includegraphics{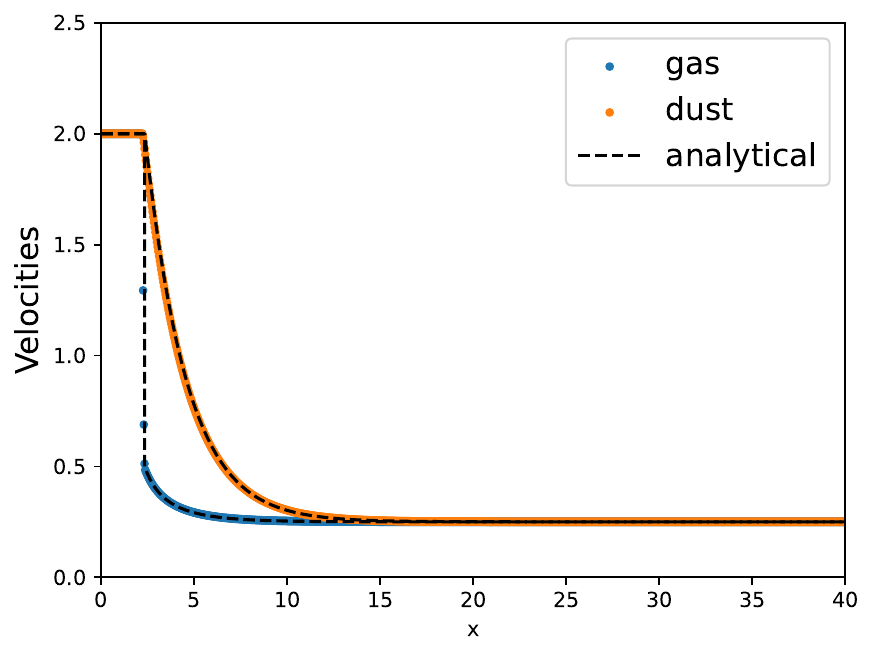}}
    \resizebox{\hsize}{!}{\includegraphics{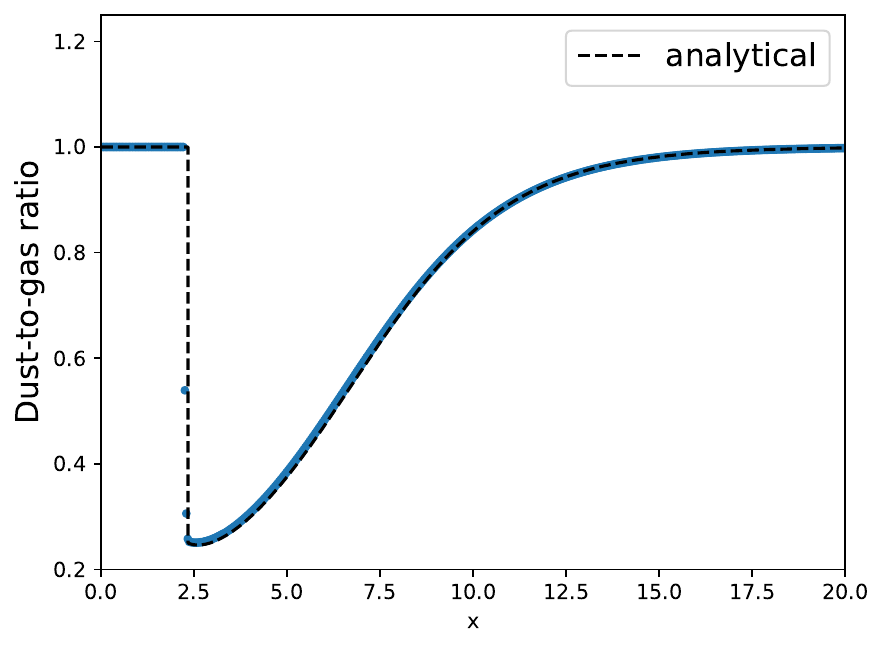}}
    \caption{Dustyshock test with the HLLgd Riemann solver, using 1024 points (level 10 of mesh refinement). Numerical solution in dots and analytical solution in black dashed lines.}
\label{fig:hlldg_dustyshock}
\end{figure}

We followed the shock setup and the solution from \citet{2019ApJS..241...25B} for one dust fluid. We set $K_d=1$ for the drag coefficient, $C_s=1$ for the sound speed, and $t=500$ for the final state. We demonstrate in Fig. \ref{fig:hlldg_dustyshock} the ability of HLLgd to capture the shock and the recoupling region.

\section{Numerical setup of protostellar collapses} \label{sec:collapse_setup}

The collapse setup follows the test of \citet{1979ApJ...234..289B}. The mass of the core is $M_0=1 M_{\odot}$. We recall, in the context of the multifluid model, the ratio between the thermal energy in the gas and the gravitational energy of the total mass
\begin{equation}
    \alpha = \frac{5}{2} \frac{1}{1+ \sum_d \theta_d} \frac{R_0}{GM_0} \frac{k_B T_g}{\mu_g m_p},
\end{equation}
where $\theta_d$ is the dust-to-gas ratio, $R_0$ is the radius of the initial core, $T_g$ is the temperature of the gas, and $\mu_g$ is the mean molecular weight. We define the ratio between the
rotational energy and the gravitational energy as
\begin{equation}
    \beta = \frac{1}{3} \frac{R_0^3 \Omega_0^2}{G M_0},
\end{equation}
where $\Omega_0$ is the angular velocity of the core. We define the mass-to-flux-to-critical-mass-to-flux ratio and, thus, the strength of the initial magnetic field, as
\begin{equation}
    \mu= \frac{M_0/\Phi_B}{(M_0/\Phi_B)_c},
\end{equation}
where $\Phi_B$ is the magnetic flux, and $(M_0/\Phi_B)_c= \frac{0.53}{3 \pi} \sqrt{5/G}$ \citep{1976ApJ...210..326M}. We set the initial conditions to $\alpha= 0.4$, $\beta=0.04$, $\mu=0.3$, $\delta \rho/\rho = 0.1$ for the initial azimuthal density perturbation, and $\phi_{\rm{mag}}=30$° for the angle between the magnetic field (z-axis in figures) and the rotation axis. The gas is coupled to the magnetic field (non-ideal MHD with ambipolar diffusion only). The reference table of ambipolar resistivities is from \citet{2016A&A...592A..18M}. The equation of state $P_g(\rho_g)$ follows the barotropic law of \citet{2013A&A...557A..90V}. We consider the Epstein regime for the drag force \citep{1924PhRv...23..710E}, for which the stopping time is recalled by Eq. \ref{eq:eipstein_stopping_time}. We use $\gamma=5/3$ for the adiabatic index of the gas. The size of the grains $s_{\rm{grain,}d}$ are typically between few nanometers to millimeters. We set the density of the grain to $\rho_{\rm{grain,}d}=1 ~\rm{g.cm^{-3}}$.

We carry out an adaptive refinement with 15 cells per local Jeans length for initially nonturbulent collapses if not specified in Sects. \ref{sec:collapse_small} and \ref{sec:collapse_large}, and with 40 cells per local Jeans length for turbulent collapses in Sect. \ref{sec:collapse_turb}. The minimum (and initial) mesh refinement level is $l_{\rm{min}}=7$ and the maximum level is $l_{\rm{max}}=14$. They correspond respectively to a cell size of $124.38$ au and $0.972$ au.

\section{Performance of the Riemann solvers for millimeter grains} \label{sec:riemann_solvers_mm}

\begin{figure*}
    \centering
    \includegraphics[width=\textwidth]{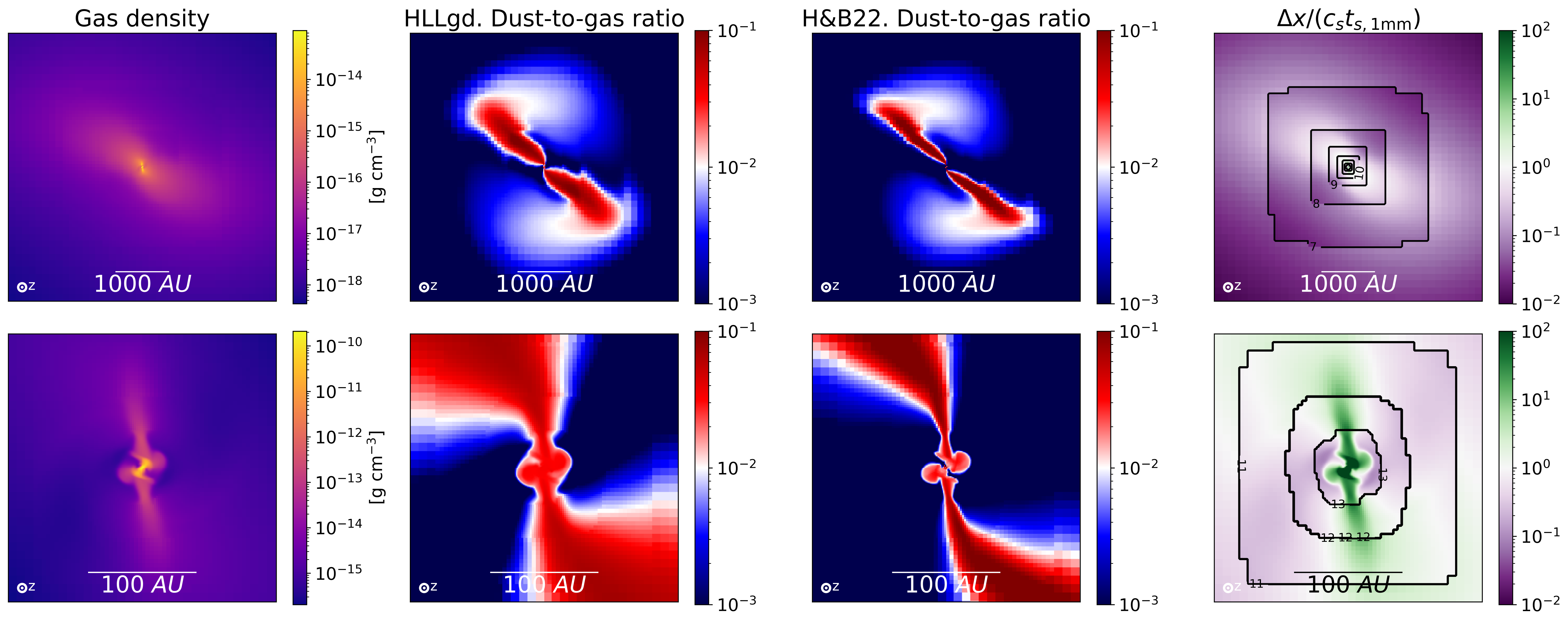}
    \caption{Comparison between two Riemann solvers for the dust fluid: HLLgd and H\&B22, the solver from \citet{2022ApJS..262...11H}. Gas density map in the first column (HLL solver) at $t= 60.6~\rm{kyr}$, dust-to-gas ratio maps from the two Riemann solvers (HLLgd in the second column and H\&B22 in the third column), and resolution of the recoupling length expressed as $\Delta x/(c_s t_{s,\rm{1mm}})$, with mesh-refinement levels indicated by contours, in the last column. Zoom-in of the collapse region (upper panels) to the disk scale (lower panels). Dust feedback has been deactivated to ease the comparison between the two solvers.
}
\label{fig:riemann_solvers_mm}
\end{figure*}

\begin{figure}
    \resizebox{\hsize}{!}{\includegraphics{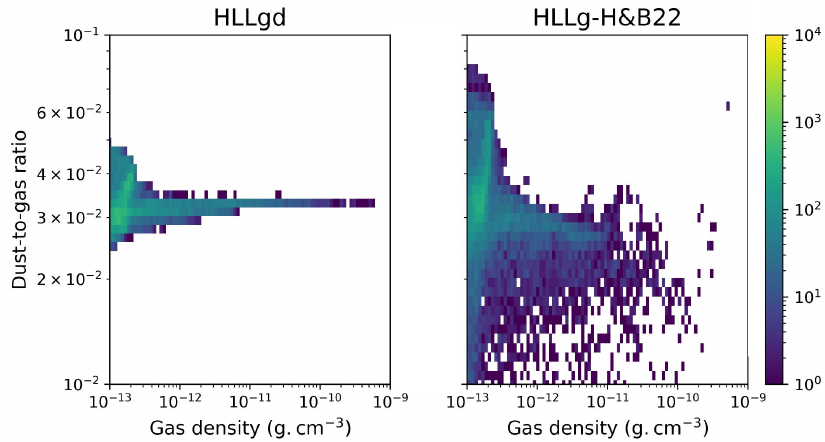}}
    \caption{Distribution of the dust-to-gas ratio as a function of the gas density (above $10^{-13}~\rm{g.cm^{-3}}$), at $t= 60.8~\rm{kyr}$. The colorbar indicates the number of cells in the bin of the histogram.}
\label{fig:hist_core_mm}
\end{figure}

In this section, we compare the performances of the HLLgd Riemann solver and the Riemann solver from \citet{2022ApJS..262...11H} denoted by H\&B22. Contrary to the rest of the paper, in this section we deactivate the dust feedback on the gas. Indeed, with such dust enrichment and velocity drift, the feedback via the gravity and the drag force is strong. We also use the same Riemann solver for the gas, that is the HLL solver, so that the gas dynamics is identical in the two simulations. It makes the comparison easier.

In Fig. \ref{fig:riemann_solvers_mm}, dust-to-gas ratio maps are different between the two solvers. Dust settles more in the pseudo-disk for the H\&B22 solver, and more generally, it produces more contrasted dust-to-gas ratio profiles, certainly because it is less diffusive than HLLgd. The main differences take place in regions where the recoupling length of the dust is not resolved (in white and in green in the last column of Fig. \ref{fig:riemann_solvers_mm}, that is in the first hydrostatic core, the disk, and the pseudo-disk). This issue worsens as the gas density increases. Indeed, we recall that $\Delta x \propto  1/\sqrt{\rho_g}$ (refinement based on the Jeans length) and $c_s t_{s,d} \propto s_{\rm{grain}}/\rho_g$, thus $\Delta x /(c_s t_{s,d}) \propto \sqrt{\rho_g}/ s_{\rm{grain}}$. Moreover, in high density regions, the velocity drift vanishes and thus the dust velocity enters in the gas wave fan, and thus Riemann solvers
differ.

More importantly, the dust recouples within the first hydrostatic core and thus the dust-to-gas ratio is homogenized. This is correctly reproduces by the HLLgd solver but not by the H\&B22 solver (Fig. \ref{fig:hist_core_mm}). Here, the H\&B22 solver lacks of robustness and can even produce negative dust densities with the chosen numerical parameters. Resolving the recoupling length within the first hydrostatic core, even for millimeter grains, requires to increase the space resolution by a factor of 100. This corresponds to a fraction of $0.01$ au, or even a fraction of the solar radius for smaller grains. Time integration of the disk formation could not be achieved with such stringent refinement.

\section{Spatial enrichment of super-micron dust grains as a function of the initial level of turbulence} \label{sec:turb_10um_100um}

\begin{figure*}
    \centering
    \includegraphics[width=\textwidth]{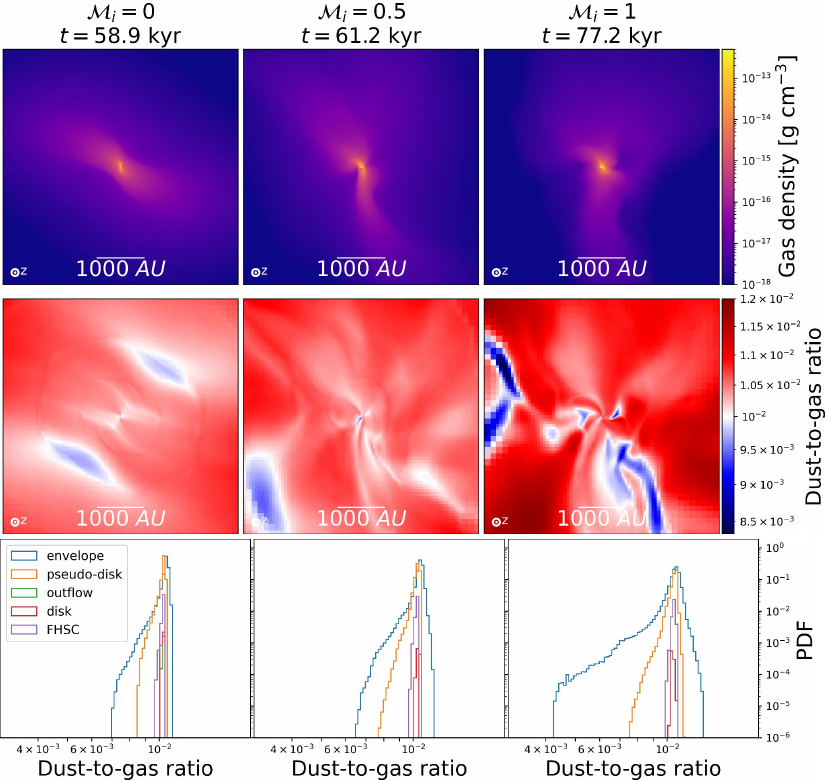}
    \caption{Same figure as Fig. \ref{fig:turbulence_map_hist_1mm}, but for 10 $\rm{\mu m}$ grains.
}
    \label{fig:turbulence_map_hist_10um}
\end{figure*}

\begin{figure*}
    \centering
    \includegraphics[width=\textwidth]{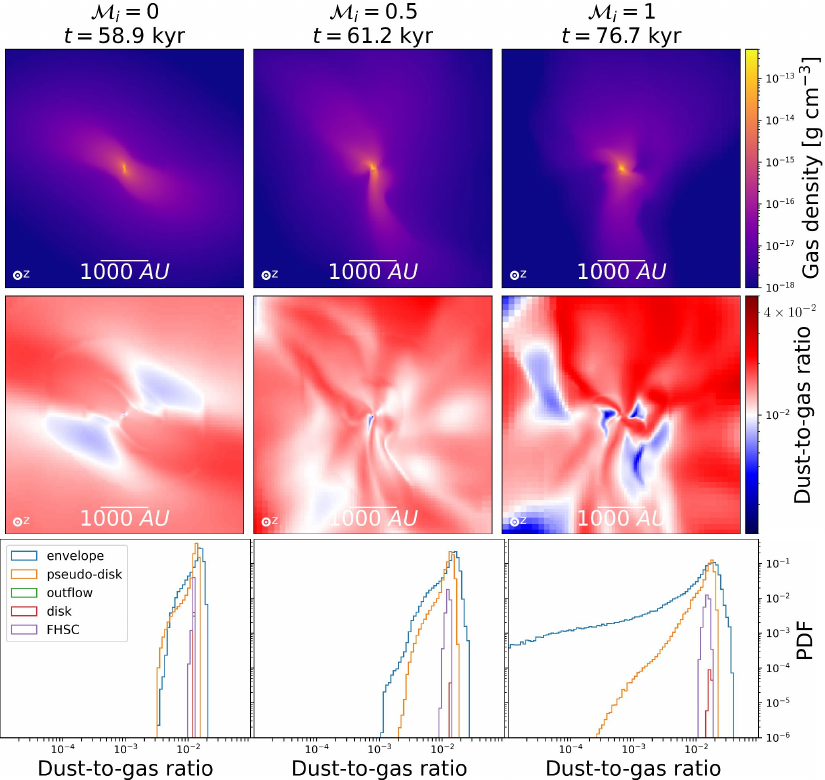}
    \caption{Same figure as Fig. \ref{fig:turbulence_map_hist_1mm}, but for 100 $\rm{\mu m}$ grains. For $\mathcal{M}_i=1$, the PDF continues to decrease for lower dust-to-gas ratio values until $\theta_d \approx 4 \times 10^{-8}$ where the PDF is about $2 \times 10^{-4}$. }
    \label{fig:turbulence_map_hist_100um}
\end{figure*}

We complete the analysis of the dust enrichment in low density regions in Sect. \ref{sec:collapse_turb} for 10 $\rm{\mu m}$ grains (Fig. \ref{fig:turbulence_map_hist_10um}) and 100 $\rm{\mu m}$ grains (Fig. \ref{fig:turbulence_map_hist_100um}). Contrary to 1 mm grains, in the envelope and the pseudo-disk, the PDF of the dust-to-gas ratio of smaller grains peaks around the enrichment within the FHSC. The dust enrichment better corresponds to the gas density profile. The main rotation plane is mostly enriched (Figs. \ref{fig:turbulence_map_hist_10um} and \ref{fig:turbulence_map_hist_100um}, face-on view). Some regions outside the rotation plane are depleted, leading to the tail of the dust-to-gas ratio PDF. The dust depletion in the envelope is more important as the initial turbulent Mach $\mathcal{M}_i$ increases. One possible explanation is that turbulence delays the FHSC formation time, providing more time for dust grains in the envelope to settle in the pseudo-disk.

\end{appendix}

\end{document}